\documentclass[aps,prd,showpacs,notitlepage,nofootinbib,superscriptaddress,floatfix,showkeys,twocolumn]{revtex4-1}

\usepackage{blindtext}
\usepackage{hyperref}
\usepackage{amsmath,amssymb}
\usepackage{float}
\usepackage{microtype}

\usepackage{graphicx}
\usepackage{bm}
\usepackage{latexsym}
\usepackage{epsfig}
\usepackage{psfrag}
\usepackage[dvipsnames]{xcolor}
\usepackage{subfigure}
\usepackage{graphicx}

\usepackage{amssymb}
\usepackage{amsmath}
\usepackage{bm}
\usepackage{latexsym}
\usepackage{epsfig}
\usepackage{psfrag}
\usepackage[normalem]{ulem}
\usepackage{textcomp}
\usepackage{color}
\usepackage{pstricks}
\usepackage[utf8]{inputenc}
\usepackage{comment}

\def\Mpl{M_{_{\mathrm{Pl}}}}
\def\f{\frac}
\def\l{\left}
\def\r{\right}
\def\d{\mathrm{d}}

\def\cP{\mathcal{P}}
\def\cL{\mathcal{L}}

\def\cJ{\mathcal{J}}
\def\cO{\mathcal{O}}

\def\tin{t_{\mathrm{in}}}
\def\tq{t_{\mathrm{q}}}
\def\Min{M_{\mathrm{in}}}

\def\Tbh{T_{_{\mathrm{BH}}}}
\def\fpbh{f_{_\mathrm{PBH}}}
\def\rh{r_{_\mathrm{H}}}

\begin{document}
\title{High-Energy Neutrino Constraints on memory burdened Bardeen and Kerr Primordial Black Holes}
\author{Md Riajul Haque}
\email{E-mail: riaj1994@sjtu.edu.cn }
\affiliation{Tsung-Dao Lee Institute $\&$ School of Physics and Astronomy, Shanghai Jiao Tong University,
Shanghai 201210, China}
\author{Suvashis Maity}
\email{E-mail: saamaity@gmail.com}
\affiliation{Indian Institute of Science Education and Research, Pune, India}


\begin{abstract}
The quantum memory burden effect can suppress the late-time Hawking evaporation of primordial black holes (PBHs), allowing PBHs below the standard evaporation threshold of approximately $10^{15}\,\mathrm{g}$ to survive until the present epoch and potentially contribute to the dark matter  abundance. We investigate primary and secondary neutrino emission from memory burdened Bardeen and Kerr PBHs, including Galactic and extragalactic fluxes and the effects of memory suppression, Bardeen regularization, and Kerr rotation. We constrain their abundance using observations from IceCube, Super-Kamiokande, and ANTARES, together with the projected sensitivities of IceCube-Gen2, GRAND200k, and Hyper-Kamiokande. We find that Bardeen regularization and Kerr rotation produce qualitatively different effects. Increasing the Bardeen parameter suppresses the Hawking temperature and neutrino flux, thereby weakening the abundance constraints. In contrast, rapidly rotating Kerr PBHs that retain a significant residual spin at the onset of the memory burdened phase can exhibit enhanced neutrino emission and substantially stronger constraints. Increasing the memory burden suppression index $k$ reduces the flux normalization but also allows lighter and hotter PBHs to survive, shifting the relevant signal toward higher energies. Consequently, HESE provides the leading sensitivity for weak suppression, whereas IceCube-Gen2 and
GRAND200k become particularly important for stronger suppression. These results demonstrate the complementarity of current and future neutrino observations in probing memory burdened PBHs and the effects of regular geometry and rotation on their evaporation.
\end{abstract}
\black
\maketitle
\section{Introduction}
\label{sec:introduction}

PBHs~\cite{Carr:1974nx} can form through the collapse of sufficiently
overdense regions in the early Universe and constitute a compelling dark matter
(DM) candidate because they are effectively cold and
noninteracting on cosmological scales~\cite{
Hawking:1971ei,Carr:1974nx,Ivanov:1994pa,Bartolo:2018evs,
Cai:2018dig,Carr:2020xqk,Jedamzik:2020ypm,Jedamzik:2020omx,
Green:2020jor,Villanueva-Domingo:2021spv,Carr:2021bzv}.
Following the detections of binary black hole mergers by the LIGO--Virgo
Collaboration~\cite{
LIGOScientific:2016aoc,LIGOScientific:2016dsl,
LIGOScientific:2016wyt,LIGOScientific:2017bnn,
LIGOScientific:2017ycc,LIGOScientific:2017vox},
considerable effort has been devoted to constraining the fraction of DM
that may reside in PBHs~\cite{
NANOGrav:2023gor,NANOGrav:2023hde,EPTA:2023sfo,EPTA:2023fyk,
Zic:2023gta,Reardon:2023gzh,Xu:2023wog,Maity:2024odg} \footnote{More recently, the LVK compact-binary-merger candidate S251112cm, which may contain a subsolar-mass component~\cite{Vieira:2026eof}, has motivated an interpretation in terms of a PBH binary~\cite{Haque:2026yum}, further highlighting the potential of gravitational-wave observations to probe the existence and abundance of PBHs.}.

PBHs lose mass through Hawking evaporation~\cite{
Hawking:1974rv,Hawking:1975vcx},
emitting Standard Model particles, including neutrinos and gamma
rays, as well as, depending on the underlying particle physics scenario,
DM, dark radiation, and high-frequency gravitational waves~\cite{
Sandick:2021gew,RiajulHaque:2023cqe,Maity:2024cpq,Green:1999yh,
Khlopov:2008qy,Belotsky:2014kca,Gondolo:2020uqv,Cheek:2021odj,
Cheek:2021cfe,Bernal:2022oha,Khlopov:2024nqp,Hooper:2019gtx,
Masina:2020xhk,Arbey:2021ysg,Barman:2024iht,Dong:2015yjs,
Sugiyama:2020roc,Inomata:2020lmk,Domenech:2021wkk,
Ireland:2023avg,Haque:2024eyh,Maity:2025ffa,Zantedeschi:2024ram,
Gross:2025hia,Bhaumik:2024qzd,Chianese:2025wrk,
Calabrese:2025sfh,Asl:2026rcm}.
These particles can affect a broad range of cosmological and
astrophysical observables. In the standard semiclassical picture, PBHs
with masses $M\lesssim10^{15}\,\mathrm{g}$ have lifetimes shorter than
the age of the Universe, whereas PBHs with
$M\lesssim10^{9}\,\mathrm{g}$ evaporate before approximately one second
and may be constrained through their effects on the light-element
abundances produced during Big Bang nucleosynthesis
(BBN)~\cite{Kohri:1999ex,Carr:2009jm,Clark:2016nst}.

During Hawking evaporation, a PBH continuously loses mass and entropy.
Because the Bekenstein--Hawking entropy is related to the logarithm of
the number of microscopic states associated with the black hole, its
decrease reflects a reduction in the number of states available to
encode information.
If evaporation is unitary, however, the information must remain encoded
in the complete quantum system.
This tension has been argued to induce a quantum backreaction known as
the memory burden effect~\cite{Dvali:2012en,Dvali:2020wft,Michel:2023ydf,Wang:2023wsm}.
This effect suppresses the evaporation rate after the black hole has
lost a sufficient fraction of its initial mass and can substantially
extend its lifetime relative to the standard Hawking prediction.
Consequently, for suitable values of the memory burden parameters, PBHs
lighter than $10^{15}\,\mathrm{g}$, which would otherwise have
evaporated by the present epoch, can survive until today and contribute
to DM.
Generally, the memory burden effect depends on two parameters: 
the remaining fraction of the initial PBH mass, $q$, at which the effect starts to dominate, 
and the memory burden parameter $k$, which determines the degree of suppression of the evaporation rate. In Sec.~\ref{sec:evap-mb}, we shall discuss these parameters in detail.
Their abundance is conventionally parametrized by
\begin{equation}
\fpbh\equiv
\frac{\rho_{\rm PBH}}{\rho_{\rm DM}},
\end{equation}
where $\rho_{\rm PBH}$ and $\rho_{\rm DM}$ are the present-day energy
densities of PBHs and total DM, respectively. Determining the
observational limits on $\fpbh$ over different mass ranges is therefore
a central objective of PBH phenomenology~\cite{
Carr:2020gox,Chen:2021ngo,Thoss:2024hsr,Chianese:2024rsn,
Montefalcone:2025akm,Ambrosone:2026djo}.
The memory burden effect opens a low-mass window that is absent in the
standard evaporation scenario, motivating a reassessment of how present
and future observations constrain the surviving PBH population.

The signal of the emitted particles from the PBH evaporation also depends on the black hole geometry. In this work, we mainly focus on the scenario where neutrinos are emitted from the evaporation. 
Regular
black hole geometries, such as the Bardeen solution, modify the
near-horizon structure and Hawking temperature through a regularization
parameter~\cite{Bardeen:1968ghaw}, while Kerr black holes introduce
rotational effects governed by their spin~\cite{
Chandrasekhar:1975zz,Chandrasekhar:1976zz,
Chandrasekhar:1977kf,TorresdelCastillo:1992zq}.
These effects can alter both the normalization and spectral shape of the
emitted neutrino flux.
Although the Hawking temperature of a Kerr black hole decreases with
increasing spin at fixed mass, the rotational chemical potential and
spin-dependent greybody factors can enhance and broaden the neutrino
spectrum when a significant residual spin survives into the
memory burdened phase.
It is therefore important to study Bardeen regularization and Kerr
rotation together with the memory burden when interpreting neutrino
limits on low-mass PBHs.

As mentioned, to explore these effects observationally, we use high-energy neutrino
measurements to constrain the abundance of memory burdened Bardeen and
Kerr PBHs. High-energy neutrinos are particularly useful for this
purpose because they can propagate over cosmological distances with
negligible attenuation and probe PBH evaporation across a broad range
of energies~\cite{
Klipfel:2025jql,Alves:2025xul,Anchordoqui:2025xug,Airoldi:2025opo}.
Related studies of neutrino emission from evaporating,
memory burdened Schwarzschild PBHs can be found in
Ref.~\cite{Chianese:2024rsn}.

We calculate the primary and secondary neutrino emission spectra and
the corresponding Galactic and extragalactic neutrino fluxes.
We compare them with the $7.5$-year IceCube high-energy starting event
(HESE) sample~\cite{IceCube:2020wum}; the upper limit on extremely
high-energy (EHE) neutrinos obtained from seven years of IceCube
data~\cite{IceCube:2016uab}; the projected sensitivities of the Giant
Radio Array for Neutrino Detection
(GRAND200k)~\cite{GRAND:2018iaj} and IceCube-Gen2~\cite{
IceCube:2019pna,IceCube-Gen2:2020qha}; the Super-Kamiokande (SK) I--IV
data set, spanning approximately 20 years~\cite{
Super-Kamiokande:2015qek}; the projected Hyper-Kamiokande sensitivity
based on the HKKM14 atmospheric-neutrino flux~\cite{
Honda:2015fha,Hyper-Kamiokande:2018ofw}; and ANTARES
data~\cite{ANTARES:2013iuz}.
This comparison allows us to determine current constraints and projected
sensitivities to $\fpbh$ throughout the surviving low-mass PBH window.

We find that increasing the Bardeen regularization parameter suppresses
the Hawking temperature and the resulting neutrino flux, thereby
weakening the limits on $\fpbh$.
By contrast, rapidly rotating Kerr PBHs can yield enhanced neutrino
emission and stronger abundance constraints when the memory burdened
phase begins before substantial spin-down, leaving a significant
residual spin. Across the surviving parameter
space, larger $k$ permits lighter and hotter PBHs to survive
until the present epoch, shifting the relevant signal toward higher
energies. Consequently, HESE provides the leading sensitivity for weak
suppression, whereas IceCube-Gen2 and GRAND200k become particularly
important for stronger suppression.

This paper is organized as follows. In Sec.~\ref{sec:evap-mb}, we review
PBH evaporation in the presence of the memory burden and introduce the
Bardeen and Kerr scenarios. In Sec.~\ref{sec:nutrino-flux}, we calculate
the neutrino flux from PBH evaporation. In
Sec.~\ref{sec:stat-analysis}, we describe the statistical methods used
to derive the limits on $\fpbh$. Finally, in Sec.~\ref{sec: result}, we
present our results and discussion.

\begin{figure}
    \centering
    \includegraphics[width=\linewidth]{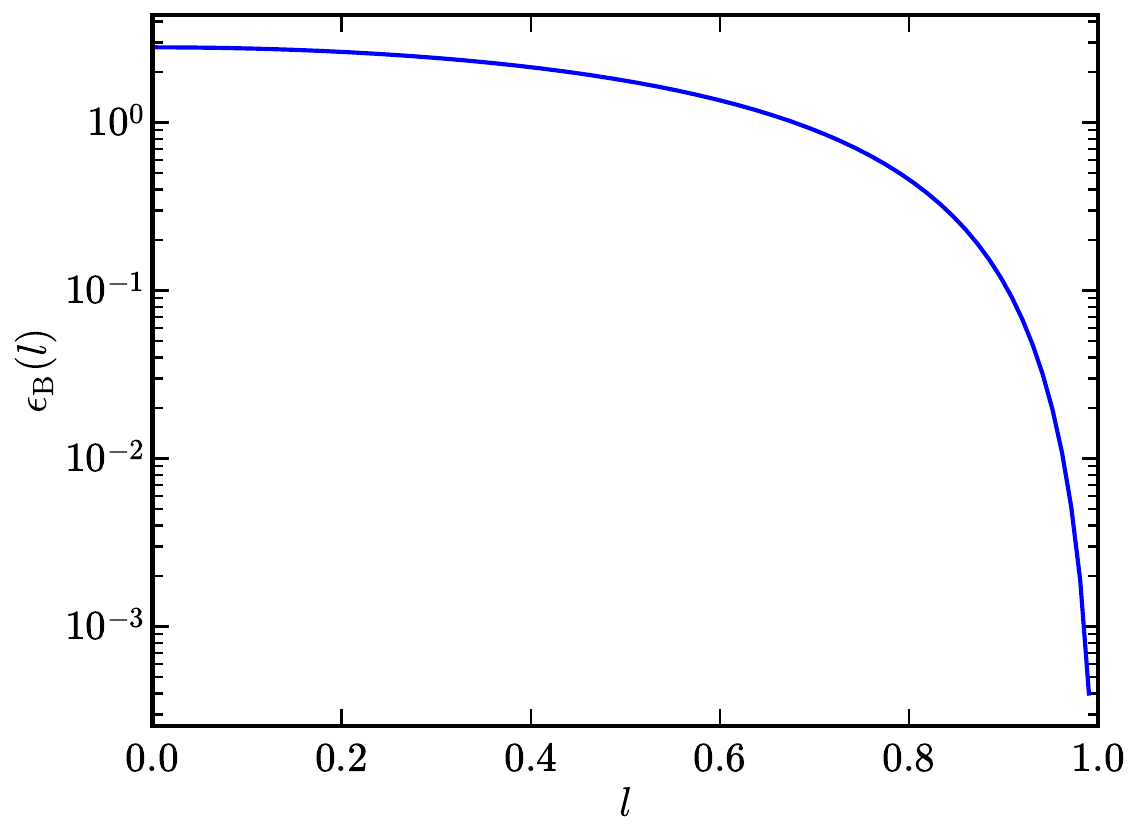}
    \caption{ The dimensionless evaporation function, obtained by integrating the energy-weighted Hawking emission spectra over all emitted particle species, $\epsilon$ for Bardeen BH is plotted here. With $l$ ranging from $0-1$, $\epsilon(l)$ varies in the range $\cO(1)-\cO(10^{-3})$.}
    \label{fig:eps-bardeen}
\end{figure}
\begin{figure*}
    \centering
    \includegraphics[width=0.49\linewidth]{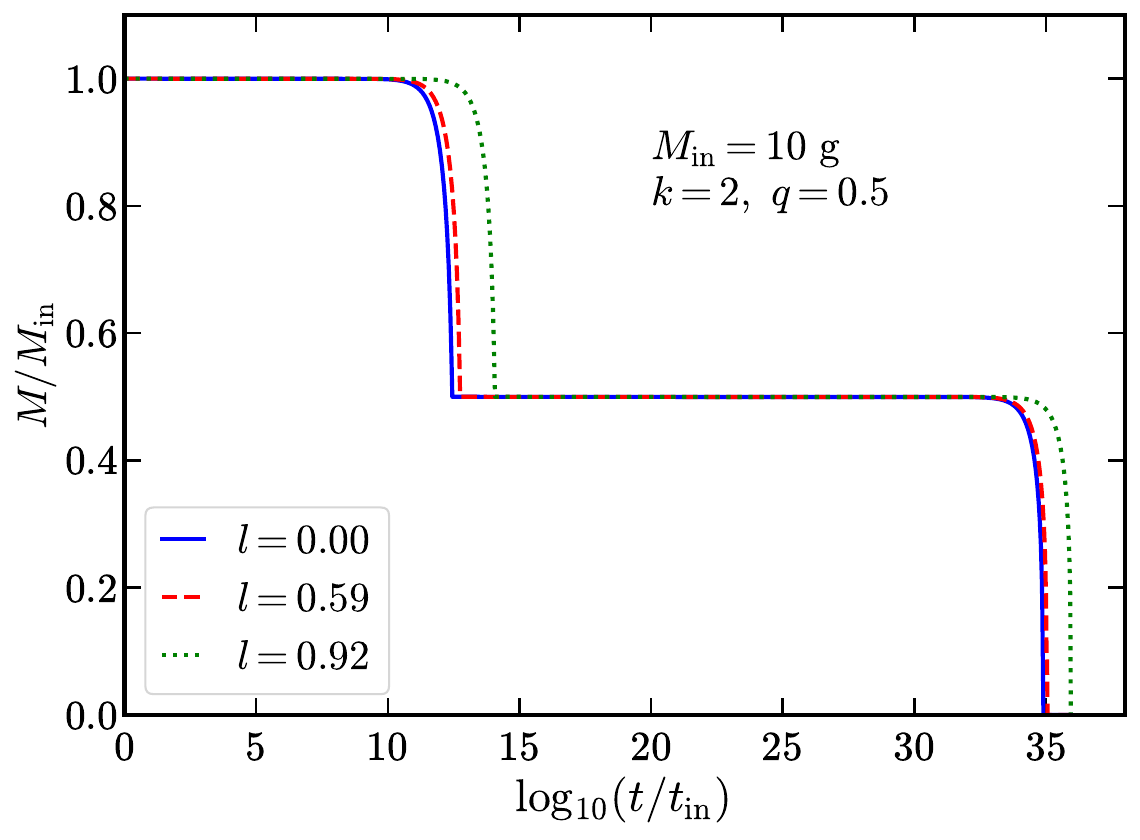}
    \includegraphics[width=0.49\linewidth]{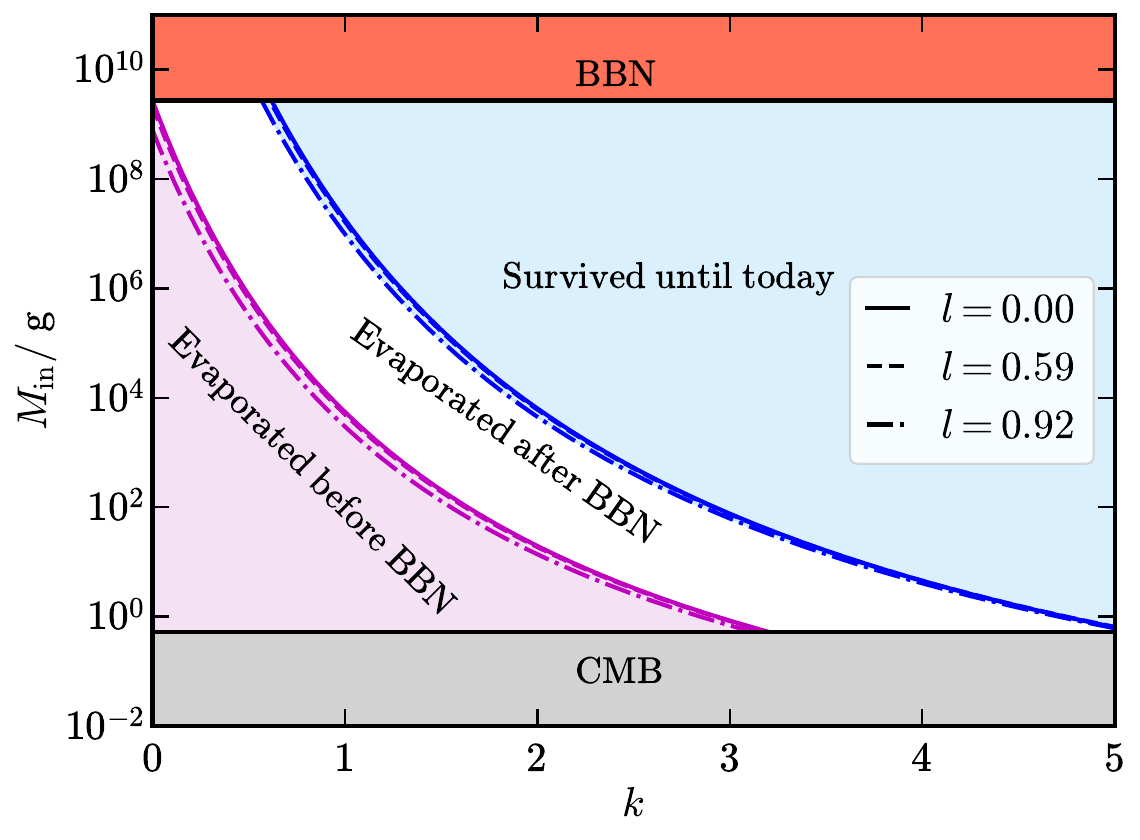}
    \caption{
    The evolution of the mass of Bardeen PBHs over time for different Bardeen parameters is plotted in the left panel. The blue solid, red dashed and green dotted lines correspond to $l=0$, $0.59$ and $0.92$, respectively.
    On the right panel, we have plotted different limits on the PBH formation mass as a function of $k$ based on whether they will survived after BBN or till today. The red line represents the $\Min$ above which the half-life of PBH will be after BBN. The blue
line represents $\Min$ above which the PBHs will remain today. The region below the magenta line represents $\Min$ for
the PBHs that evaporated before BBN. 
The black line correspond to the minimum allowed value of
the formation mass, $\Min$, which is calculated using the maximum allowed value of the energy scale of inflation, i.e.
$H_{\rm inf} \sim 5 \times 10^{13}$ GeV, set by the upper bound of the tensor-to scalar ratio $r \leq 0.036$ from CMB data~\cite{Planck:2018vyg,Planck:2018jri}.}
    \label{fig:m-k-bardeen}
\end{figure*}

\begin{figure*}
    \centering
    \includegraphics[width=0.49\linewidth]{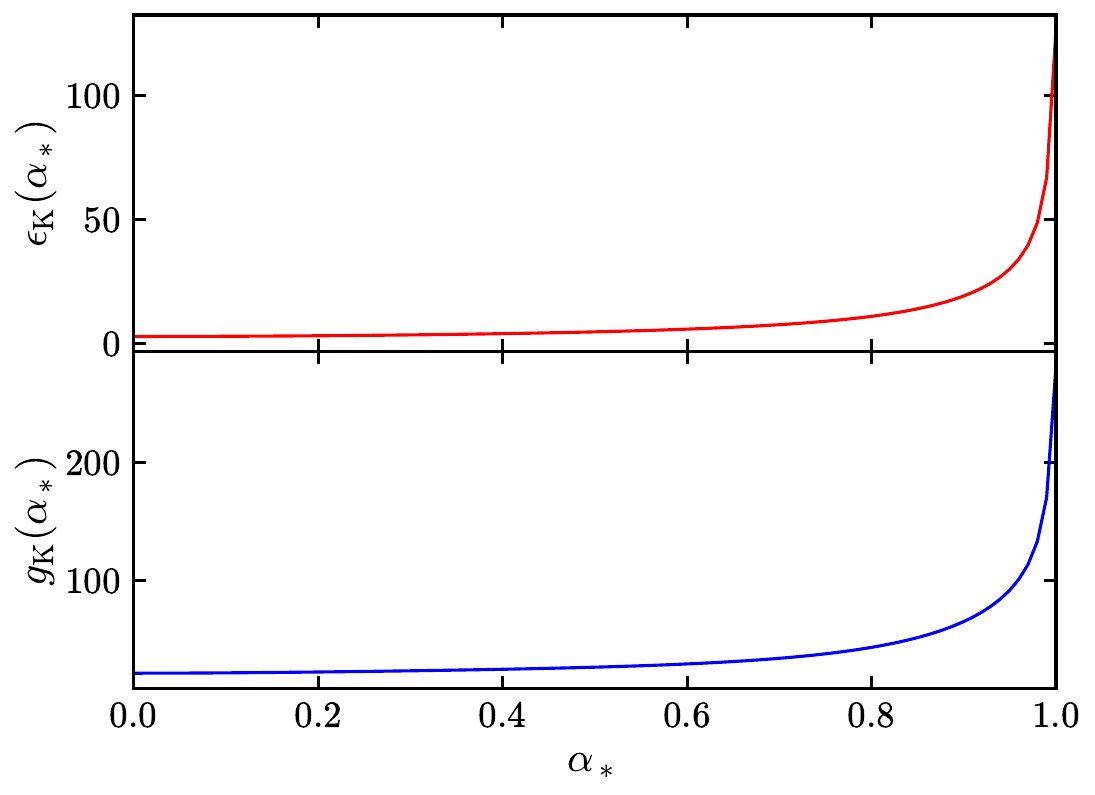}
    \includegraphics[width=0.49\linewidth]{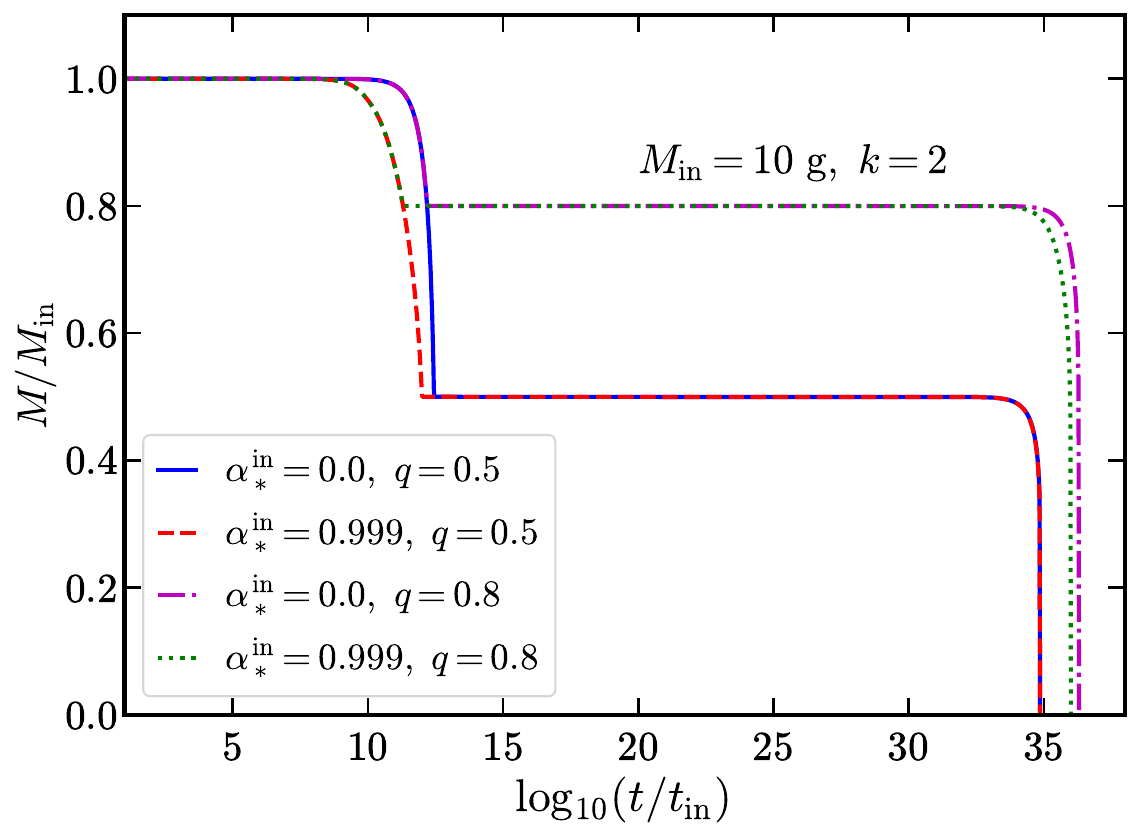}
    \includegraphics[width=0.49\linewidth]{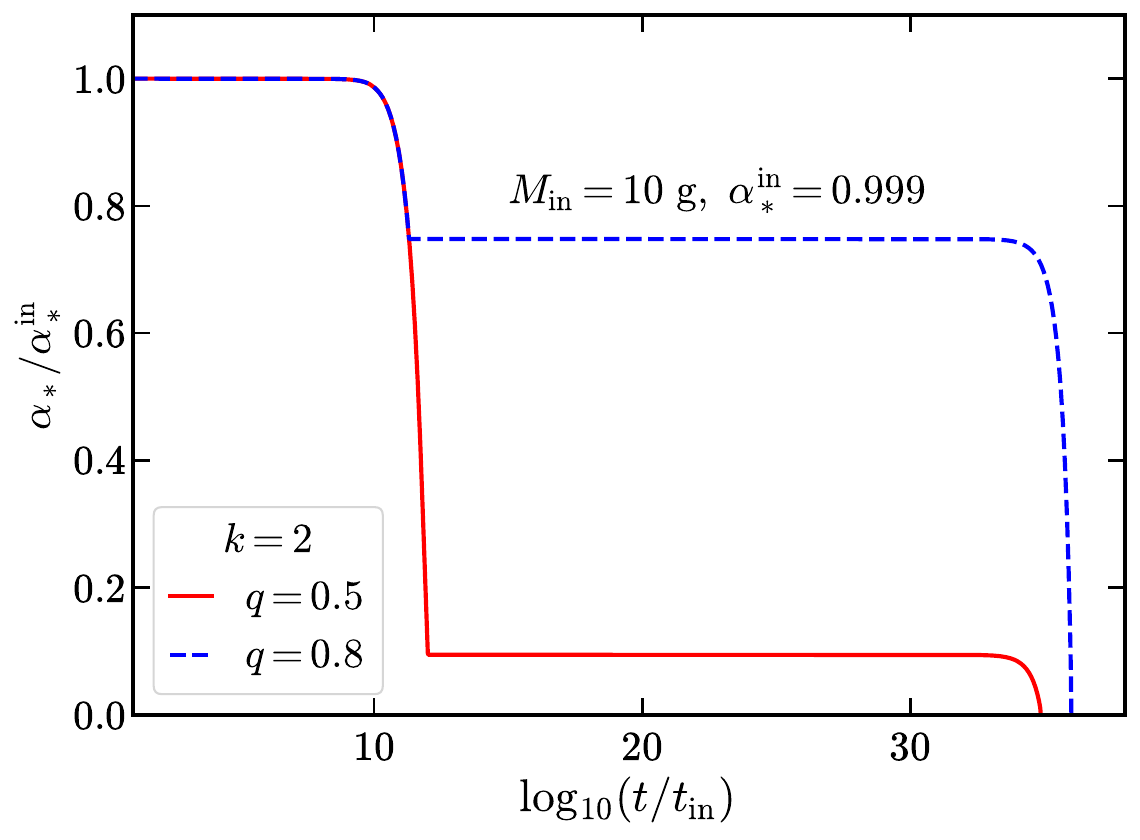}
    \includegraphics[width=0.49\linewidth]{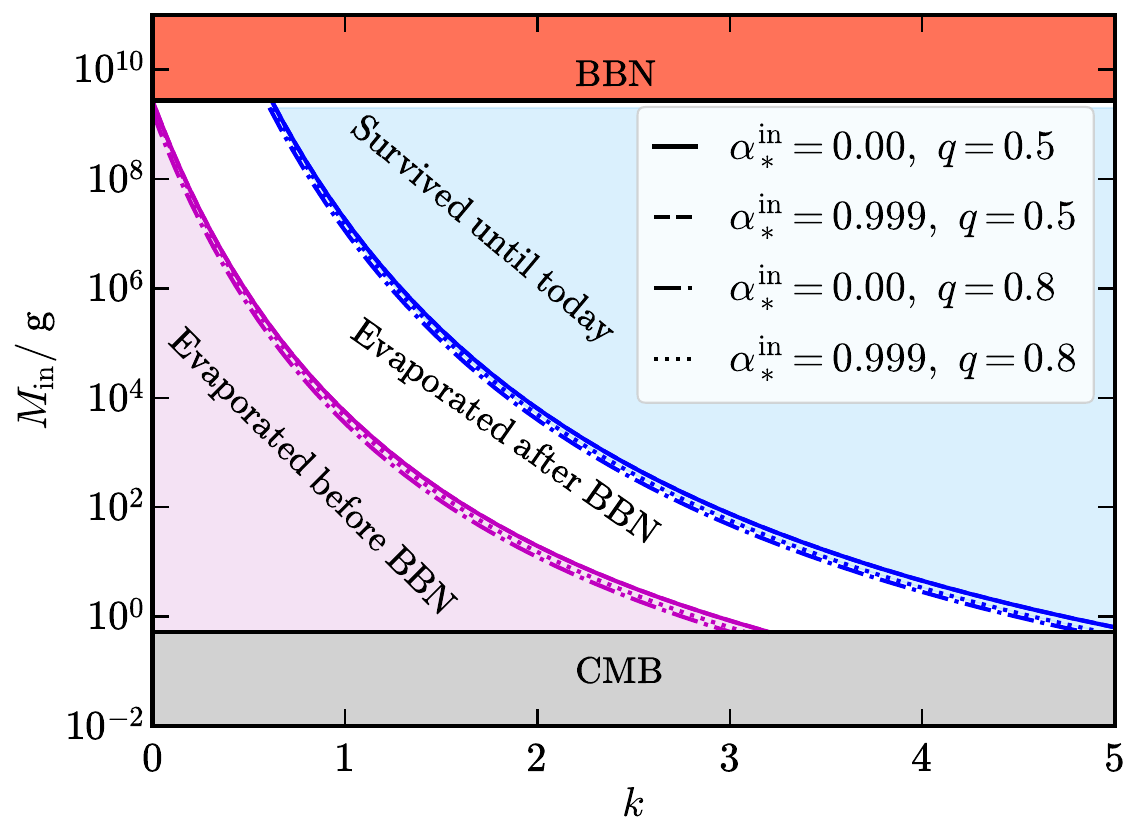}
    \caption{The dimensionless Page functions governing the loss of energy and angular momentum i.e. $\epsilon(M,\alpha_\ast)$ and $g(M,\alpha_\ast)$ are plotted as a function of the spin parameter $\alpha_\ast$ on the top left panel in red and blue respectively.
    The evolution of the PBH mass for memory burden parameter $k=2$ is plotted on the top right panel. The blue, red dashed, magenta dot-dashed and green dotted lines correspond to the parameter value $(\alpha_\ast^{\rm in},q)$ as $(0.0,0.5)$, $(0.999,0.5)$, $(0.0,0.8)$ and $(0.999,0.8)$, respectively.
    The evolution of PBH spin for the PBH with initial spin $\alpha_\ast^{\rm in}=0.999$ and $k=2$ is plotted as function of time on the bottom left panel with $q=0.5$ in red and $q=0.8$ in blue dashed lines respectively.
     Different limits on the formation mass as a function of $k$ based on whether they will survived after BBN or till today is plotted on the bottom right panel.  
The solid, dashed dot-dashed and dotted lines correspond to the parameter value $(\alpha_\ast^{\rm in},q)$ as $(0.0,0.5)$, $(0.999,0.5)$, $(0.0,0.8)$ and $(0.999,0.8)$, respectively.
The Blue, magenta, red and black lines correspond to the parameters similar to Fig.~\ref{fig:m-k-bardeen}.
    }
    \label{fig:mass-spin-evol}
\end{figure*}

In this paper, we shall work with natural units such that $\hbar=c=1$, and the reduced 
Planck mass is $\Mpl=\l(8\,\pi\, G\r)^{-1/2} \simeq 2.4 \times10^{18}\,
\mathrm{GeV}$.
The cosmic time, conformal time and redshift are denoted by $t$, $\eta$ and $z$, respectively.

\section{Evaporation of PBHs}
\label{sec:evap-mb}

PBHs are assumed to form in the early Universe through the collapse of overdense regions. For a PBH with initial mass $\Min$, the formation time is $\tin=\Min/(8\pi\gamma\Mpl^2)$, where $\gamma=0.2$ is the collapse efficiency for PBHs formed during radiation domination~\cite{Carr:1974nx}. The subsequent evolution depends on the BH geometry. In this work, we consider the regular, nonrotating Bardeen geometry and the rotating Kerr geometry.

We first consider the Bardeen BH, a well-known example of a regular BH. Its line element is~\cite{Bardeen:1968ghaw}
\begin{align}
    \d s^2
    =
    -f(r)\d t^2
    +\f{\d r^2}{f(r)}
    +r^2\d\Omega^2,
\end{align}
where
\begin{align}
    f(r)
    =
    1-\f{M}{4\pi\Mpl^2}
    \f{r^2}{(r^2+\ell^2)^{3/2}},
\end{align}
and $\ell$ is the regularization scale. In the limit $\ell=0$, the metric reduces to the Schwarzschild solution. Unlike the Schwarzschild geometry, however, the Bardeen geometry is regular at the center, $r=0$.

For this geometry to possess an event horizon, the regularization scale must satisfy $\ell\leq\ell_{\rm crit}$, where $\ell_{\rm crit}=M/(6\sqrt{3}\,\pi\Mpl^2)$. We define the dimensionless Bardeen parameter as
$l\equiv\ell/\ell_{\rm crit}=6\sqrt{3}\,\pi(\Mpl^2/M)\ell$.
We restrict our analysis to nonextremal Bardeen BHs, for which $l\in[0,1)$; the limiting value $l=1$ corresponds to the extremal configuration. The metric function then becomes
\begin{align}
    f(r)
    =
    1-\f{Mr^2}{4\pi\Mpl^2}
    \left(
        r^2+\f{M^2l^2}{108\pi^2\Mpl^4}
    \right)^{-3/2}.
\end{align}
The horizon radius $\rh$ is determined by $f(\rh)=0$, which gives
\begin{align}
    \left(
        \rh^2+\f{M^2l^2}{108\pi^2\Mpl^4}
    \right)^{3/2}
    =
    \f{M\rh^2}{4\pi\Mpl^2}.
\end{align}
For specified values of $M$ and $l$, this equation is solved numerically for the outer horizon radius.

The Hawking temperature and Bekenstein--Hawking entropy of a Bardeen black hole are
\begin{align}
    \Tbh^{\rm Bardeen}
    &=
    \f{1}{4\pi\rh}
    \f{
        \rh^2-\f{M^2l^2}{54\pi^2\Mpl^4}
    }{
        \rh^2+\f{M^2l^2}{108\pi^2\Mpl^4}
    },
    \\
    S_{\rm Bardeen}
    &=
    8\pi^2\Mpl^2\rh^2.
    \label{eq:s-bardeen}
\end{align}
Throughout our analysis of Bardeen PBH, the dimensionless parameter $l$ is held fixed during evaporation. Under this prescription, $\rh\propto M$, so that $S_{\rm Bardeen}\propto M^2$ and $\Tbh^{\rm Bardeen}\propto M^{-1}$. Thus, the entropy decreases and the temperature increases as the PBH loses mass. At fixed $M$, increasing $l$ reduces both the horizon radius and the surface gravity, thereby lowering the entropy and Hawking temperature. This qualification is important because fixing the dimensionful scale $\ell$, rather than $l$, would lead to a different late-time evolution toward the extremal limit.

Next, we consider spinning PBHs described by the Kerr geometry. In Boyer--Lindquist coordinates $(t,r,\theta,\phi)$, we define
\begin{align}
    \Sigma&=r^2+a^2\cos^2\theta,
    &
    \Delta&=r^2-\f{Mr}{4\pi\Mpl^2}+a^2,
    \nonumber\\
    A&=r^2+a^2,
    &
    B&=\f{Mr}{4\pi\Mpl^2\Sigma}.
\end{align}
The Kerr line element can then be written compactly as~\cite{Chandrasekhar:1975zz,Chandrasekhar:1976zz,Chandrasekhar:1977kf,TorresdelCastillo:1992zq}
\begin{align}
    \d s^2={}&
    -(1-B)\d t^2
    -2Ba\sin^2\theta\,\d t\,\d\phi
    +\f{\Sigma}{\Delta}\d r^2
    +\Sigma\,\d\theta^2
    \nonumber\\
    &+
    \left(
        A+Ba^2\sin^2\theta
    \right)
    \sin^2\theta\,\d\phi^2.
\end{align}
Here, $J$ is the PBH angular momentum and $a=J/M$ is the dimensionful Kerr rotation parameter. The dimensionless spin parameter is
\begin{align}
    \alpha_\ast
    \equiv
    8\pi\Mpl^2\f{a}{M}
    =
    8\pi\Mpl^2\f{J}{M^2},
    \qquad
    |\alpha_\ast|\leq1.
\end{align}
The outer and inner horizon radii follow from $\Delta=0$:
\begin{align}
    r_\pm
    =
    \f{M}{8\pi\Mpl^2}
    \left(
        1\pm\sqrt{1-\alpha_\ast^2}
    \right).
\end{align}
The Hawking temperature and Bekenstein--Hawking entropy are
\begin{align}
    \Tbh^{\rm Kerr}
    &=
    \f{2\Mpl^2}{M}
    \f{\sqrt{1-\alpha_\ast^2}}
    {1+\sqrt{1-\alpha_\ast^2}},
    \nonumber\\
    S_{\rm Kerr}
    &=
    \f{1}{4}
    \left(
        \f{M}{\Mpl}
    \right)^2
    \left(
        1+\sqrt{1-\alpha_\ast^2}
    \right).
    \label{eq:temp-entrop}
\end{align}
In the nonrotating limit, $\alpha_\ast=0$, the Kerr geometry reduces to the Schwarzschild geometry, for which $\Tbh=\Mpl^2/M$ and $S=(1/2)(M/\Mpl)^2$.
At fixed mass, both the temperature and entropy decrease monotonically as $\alpha_\ast$ increases, with the temperature vanishing in the extremal limit. Nevertheless, a lower Kerr temperature does not necessarily imply weaker particle emission, because rotation also modifies the thermal factor and greybody coefficients and can enhance the emission of co-rotating modes.

Having specified the geometries and their thermodynamic properties, we now describe PBH evaporation in the presence of the memory burden. For compactness, we introduce
\begin{align}
    X\in\{{\rm B},{\rm K}\},
    \qquad
    \lambda_{\rm B}=l,
    \qquad
    \lambda_{\rm K}=\alpha_\ast.
\end{align}
The mass-loss rate is given by
\begin{align}
    \f{\d M}{\d t}&=
    -\epsilon_X(M,\lambda_X)
    \f{\Mpl^4}{M^2}
    S_X^{-k\Theta(q\Min-M)},
    \label{eq:dmdt}
\end{align}
where $\Theta$ is the Heaviside step function. When $M>q\Min$, one has $\Theta(q\Min-M)=0$, and the PBH undergoes standard Hawking evaporation. Once $M$ reaches $q\Min$, the memory burden becomes relevant and suppresses the mass-loss rate by $S_X^{-k}$. The parameter $q$ determines the onset of this phase, while $k>0$ controls the strength of the suppression. Since the entropy of a semiclassical PBH is large, even a moderate value of $k$ can substantially extend its lifetime.

For either geometry, the dimensionless evaporation function is defined by
\begin{align}
    \epsilon_X(M,\lambda_X)=
    \f{M^2}{\Mpl^4}
    \int_0^\infty \d E\,E
    \sum_j
    \f{\d^2N_j}
    {\d t\,\d E}
    \left(E;M,\lambda_X\right),
    \label{eq:epsilon-general}
\end{align}
where the sum runs over all primary particle species and internal degrees of freedom emitted by the PBH. The integral is the total Hawking power, while the prefactor makes $\epsilon_X$ dimensionless and ensures consistency with Eq.~\eqref{eq:dmdt}. The dependence on $M$ accounts primarily for particle-mass thresholds, whereas the geometric dependence enters through the Hawking temperature and greybody factors. The Bardeen and Kerr functions are obtained as $\epsilon_{\rm B}(M,l)$ and $\epsilon_{\rm K}(M,\alpha_\ast)$, respectively.

For a Kerr PBH, we also define the dimensionless angular-momentum Page function
\begin{align}
    g_{\rm K}(M,\alpha_\ast)
    &\equiv
    -\f{8\pi M}{\alpha_\ast\Mpl^2}
    \left(
        \f{\d J}{\d t}
    \right)_{\rm H}
    \nonumber\\
    &=
    \f{8\pi M}{\alpha_\ast\Mpl^2}
    \int_0^\infty \d E
    \sum_{j,\tilde{\ell},m}
    m\,
    \f{\d^2N_{j\tilde{\ell}m}}
    {\d t\,\d E},
    \label{eq:g-kerr}
\end{align}
where $(\d J/\d t)_{\rm H}$ is the semiclassical Hawking angular-momentum loss rate, $\tilde{\ell}$ and $m$ are the angular-mode quantum numbers, and the modal spectrum includes the particle multiplicities. Eqs.~\eqref{eq:epsilon-general} and~\eqref{eq:g-kerr} use the same normalization convention. In the phenomenological model adopted here, the memory burden suppresses the energy and angular-momentum fluxes by the same factor $S_{\rm K}^{-k}$.

For Bardeen PBHs, the dependence of $\epsilon_{\rm B}$ on $l$ is shown in the Fig.~\ref{fig:eps-bardeen}. Increasing $l$ lowers the Hawking temperature and modifies the greybody factors, reducing the total emitted power and hence $\epsilon_{\rm B}$. In the limit $l=0$, the Bardeen geometry and its evaporation function reduce to their Schwarzschild counterparts.

For both geometries, the entropy can be written as
\begin{align}
    S_X
    =
    C_X
    \left(
        \f{M}{\Mpl}
    \right)^2,
    \qquad
    X\in\{{\rm B},{\rm K}\},
\end{align}
with
\begin{align}
    C_{\rm B}(l)
    &=
    8\pi^2
    \f{\Mpl^4\rh^2}{M^2},
    \nonumber\\
    C_{\rm K}(\alpha_\ast)
    &=
    \f{1}{4}
    \left(
        1+\sqrt{1-\alpha_\ast^2}
    \right).
\end{align}
For fixed $l$, the ratio $\Mpl^2\rh/M$ is constant, and therefore $C_{\rm B}$ depends only on $l$.

For a Kerr PBH, we denote the mass and residual spin at the onset of the memory burdened phase by
\begin{align}
    M_q=q\Min,
    \qquad
    \alpha_{\ast}^{\rm q}
    =
    8\pi\Mpl^2
    \f{J_q}{(q\Min)^2},
\end{align}
where $J_q$ is the angular momentum remaining at $M=q\Min$. Hence, $\alpha_{\ast}^{\rm q}$ depends on both $q$ and the initial spin $\alpha_{\ast}^{\rm in}$.

We next derive analytical lifetime estimates under the constant-parameter approximation. For Bardeen PBHs, $l$ and $\epsilon_{\rm B}(l)$ are treated as constant. For Kerr PBHs, the spin is fixed at $\alpha_{\ast}^{\rm q}$, and both $C_{\rm K}$ and $\epsilon_{\rm K}$ are evaluated at this residual spin during the memory burdened phase. Under these assumptions, the duration of the burdened phase is
\begin{align}
    t_{{\rm m},X}^{\,k}
    &\simeq
    \f{[C_X]^k}
    {(3+2k)\epsilon_X\Mpl}
    \left(
        \f{q\Min}{\Mpl}
    \right)^{3+2k},
    \label{eq:tm-general}
\end{align}
where $t_{{\rm m},X}^{\,k}$ is measured from $M=q\Min$ to the formal endpoint of evaporation within the adopted mass-loss model. Explicitly,
\begin{align}
    t_{{\rm m},{\rm B}}^{\,k}(l)
    &\simeq
    \f{[C_{\rm B}(l)]^k}
    {(3+2k)\epsilon_{\rm B}(l)\Mpl}
    \left(
        \f{q\Min}{\Mpl}
    \right)^{3+2k},
    \nonumber\\
    t_{{\rm m},{\rm K}}^{\,k}
    \left(\alpha_{\ast}^{\rm q}\right)
    &\simeq
    \f{
        [C_{\rm K}(\alpha_{\ast}^{\rm q})]^k
    }{
        (3+2k)
        \epsilon_{\rm K}(\alpha_{\ast}^{\rm q})
        \Mpl
    }
    \left(
        \f{q\Min}{\Mpl}
    \right)^{3+2k},
\end{align}
where
\begin{align}
    C_{\rm K}\left(\alpha_{\ast}^{\rm q}\right)
    =
    \f{1}{4}
    \left[
        1+
        \sqrt{
            1-\left(\alpha_{\ast}^{\rm q}\right)^2
        }
    \right].
\end{align}

Let $\tq$ denote the cosmic time at which the PBH reaches $q\Min$. Approximating the evaporation function as constant during the semiclassical phase gives
\begin{align}
    \tq-\tin
    &\simeq
    \f{1-q^3}
    {3\,\epsilon_{\rm B}(l)\Mpl}
    \left(
        \f{\Min}{\Mpl}
    \right)^3
\end{align}
for a Bardeen PBH. For a Kerr PBH, the spin changes most rapidly toward the end of the semiclassical evolution. A useful analytical estimate is therefore obtained by evaluating the evaporation function at the initial spin:
\begin{align}
    \tq-\tin
    &\simeq
    \f{1-q^3}
    {3\,\epsilon_{\rm K}
        \left(\alpha_{\ast}^{\rm in}\right)\Mpl}
    \left(
        \f{\Min}{\Mpl}
    \right)^3.
\end{align}
The residual spin $\alpha_{\ast}^{\rm q}$, rather than $\alpha_{\ast}^{\rm in}$, is used for the subsequent memory burdened phase.

The total lifetime measured from formation is
\begin{align}
    t_{{\rm ev},X}^{\,k}
    =
    (\tq-\tin)
    +t_{{\rm m},X}^{\,k}.
\end{align}
For the parameter region considered in this work, the burdened phase dominates the lifetime:
\begin{align}
    t_{{\rm m},X}^{\,k}
    \gg
    \tq-\tin,
    \qquad
    t_{{\rm ev},X}^{\,k}
    \simeq
    t_{{\rm m},X}^{\,k}.
\end{align}
The corresponding approximate mass evolution is
\begin{align}
    M(t)
    &\simeq
    \begin{cases}
        \displaystyle
        \Min
        \left[
            1-\f{t-\tin}{t_{{\rm ev},X}^{\,0}}
        \right]^{1/3},
        &
        \tin\leq t<\tq,
        \\[2mm]
        \displaystyle
        q\Min
        \left[
            1-\f{t-\tq}{t_{{\rm m},X}^{\,k}}
        \right]^{1/(3+2k)},
        &
        \tq\leq t<\tq+t_{{\rm m},X}^{\,k}.
    \end{cases}
\end{align}
The semiclassical evaporation times are
\begin{align}
    t_{{\rm ev},{\rm B}}^{\,0}(l)
    &\simeq
    \f{1}
    {3\,\epsilon_{\rm B}(l)\Mpl}
    \left(
        \f{\Min}{\Mpl}
    \right)^3,
    \nonumber\\
    t_{{\rm ev},{\rm K}}^{\,0}
    \left(\alpha_{\ast}^{\rm in}\right)
    &\simeq
    \f{1}
    {
        3\,\epsilon_{\rm K}
        \left(\alpha_{\ast}^{\rm in}\right)\Mpl
    }
    \left(
        \f{\Min}{\Mpl}
    \right)^3.
\end{align}

The Schwarzschild result is recovered from either geometry. For $l=0$ and $\alpha_\ast=0$, one has
\begin{align}
    C_{\rm B}(0)
    =
    C_{\rm K}(0)
    =
    \f{1}{2},
    \quad
    \epsilon_{\rm B}(M,0)
    =
    \epsilon_{\rm K}(M,0)
    =
    \epsilon_{\rm Sch}(M).
\end{align}
The burdened-phase lifetime then becomes
\begin{align}
    t_{{\rm m},{\rm Sch}}^{\,k}
    \simeq
    \left[
        2^k(3+2k)\epsilon_{\rm Sch}\Mpl
    \right]^{-1}
    \left(
        \f{q\Min}{\Mpl}
    \right)^{3+2k},
\end{align}
which is the standard memory burdened Schwarzschild result under the constant-evaporation-function approximation.

The analytical expressions above fix the relevant geometric parameters within each phase. In the numerical Kerr analysis, however, the mass and spin are evolved simultaneously. Using Eqs.~\eqref{eq:epsilon-general} and~\eqref{eq:g-kerr}, the coupled equations are
\begin{align}
    \f{\d M}{\d t}
    &=
    -\epsilon_{\rm K}(M,\alpha_\ast)
    \f{\Mpl^4}{M^2}
    S_{\rm K}^{-k\Theta(q\Min-M)},
    \nonumber\\
    \f{\d\alpha_\ast}{\d t}
    &=
    \f{\alpha_\ast}{M}
    \left[
        \f{g_{\rm K}(M,\alpha_\ast)}
        {\epsilon_{\rm K}(M,\alpha_\ast)}
        -2
    \right]
    \f{\d M}{\d t}
    \nonumber\\
    &=
    \alpha_\ast
    \f{\Mpl^4}{M^3}
    \left[
        2\epsilon_{\rm K}(M,\alpha_\ast)
        -g_{\rm K}(M,\alpha_\ast)
    \right]
    S_{\rm K}^{-k\Theta(q\Min-M)}.
    \label{eq:kerr-coupled-evolution}
\end{align}
The initial and transition conditions are
\begin{align}
    \alpha_\ast(\tin)
    =
    \alpha_{\ast}^{\rm in},
    \qquad
    \alpha_\ast(\tq)
    =
    \alpha_{\ast}^{\rm q}.
\end{align}
Equation~\eqref{eq:kerr-coupled-evolution} is used for the numerical Kerr evolution and the results shown in Fig.~\ref{fig:mass-spin-evol}; the preceding lifetime and mass-evolution formulas are analytical estimates.

In the left panel of Fig.~\ref{fig:m-k-bardeen}, we show the mass evolution of Bardeen PBH for $k=2$ and $\Min=10~{\rm g}$. The blue solid, red dashed, and green dotted curves correspond to $l=0$, $l=0.59$ $(\ell=0.25\,\rh)$, and $l=0.92$ $(\ell=0.5\,\rh)$, respectively. We use these representative values throughout the analysis. 
As $l$ increases, $\epsilon_{\rm B}(l)$ decreases, slowing the mass loss and extending the PBH lifetime.

The right panel of Fig.~\ref{fig:m-k-bardeen} shows the corresponding boundaries in the $(k,\Min)$ plane. The solid, dashed, and dash-dotted curves represent $l=0$, $l=0.59$, and $l=0.92$, respectively. Because a larger $l$ extends the lifetime, a smaller initial mass is required for the PBH to evaporate within a fixed cosmic time. The blue region identifies PBHs that survive until the present epoch but complete their initial semiclassical evaporation phase before BBN.

The top-right and bottom-left panels of Fig.~\ref{fig:mass-spin-evol} show the mass and spin evolution of Kerr PBH, respectively, for $k=2$ and $\Min=10~{\rm g}$. We consider $\alpha_{\ast}^{\rm in}=0$ and $0.999$, together with $q=0.5$ and $0.8$. In the top-right panel, the blue solid, red dashed, magenta dot-dashed, and green dotted curves correspond to
   $ (\alpha_{\ast}^{\rm in},q)
    =
    (0,0.5),\,
    (0.999,0.5),\,
    (0,0.8),\,
    (0.999,0.8),$
respectively. In the bottom-left panel, the red solid and blue dashed curves correspond to $q=0.5$ and $q=0.8$, respectively, for $\alpha_{\ast}^{\rm in}=0.999$. 

A larger $q$ activates the memory burden while more of the initial mass remains and therefore extends the PBH lifetime. Whereas, rapid rotation increases the Kerr evaporation function over the parameter range considered, enhancing the mass-loss rate and shortening the lifetime. The spin evolution is governed by the competition between energy and angular-momentum loss encoded in $2\epsilon_{\rm K}-g_{\rm K}$. For $\alpha_{\ast}^{\rm in}=0.999$, the numerical evolution gives approximately $\alpha_{\ast}^{\rm q}\simeq0.75\,\alpha_{\ast}^{\rm in}$ for $q=0.8$ and $\alpha_{\ast}^{\rm q}\simeq0.1\,\alpha_{\ast}^{\rm in}$ for $q=0.5$. Thus, an earlier transition preserves a substantially larger residual spin. Finally, the bottom-right panel of Fig.~\ref{fig:mass-spin-evol} shows the corresponding boundaries in the $(k,\Min)$ plane. Since increasing $q$ extends the lifetime, the largest initial mass that can evaporate within a fixed cosmic time decreases with $q$. Whereas, increasing $\alpha_{\ast}^{\rm in}$ enhances the evaporation rate and shifts this boundary in the opposite direction.


\begin{figure*}
    \centering
    \includegraphics[width=.9\linewidth]{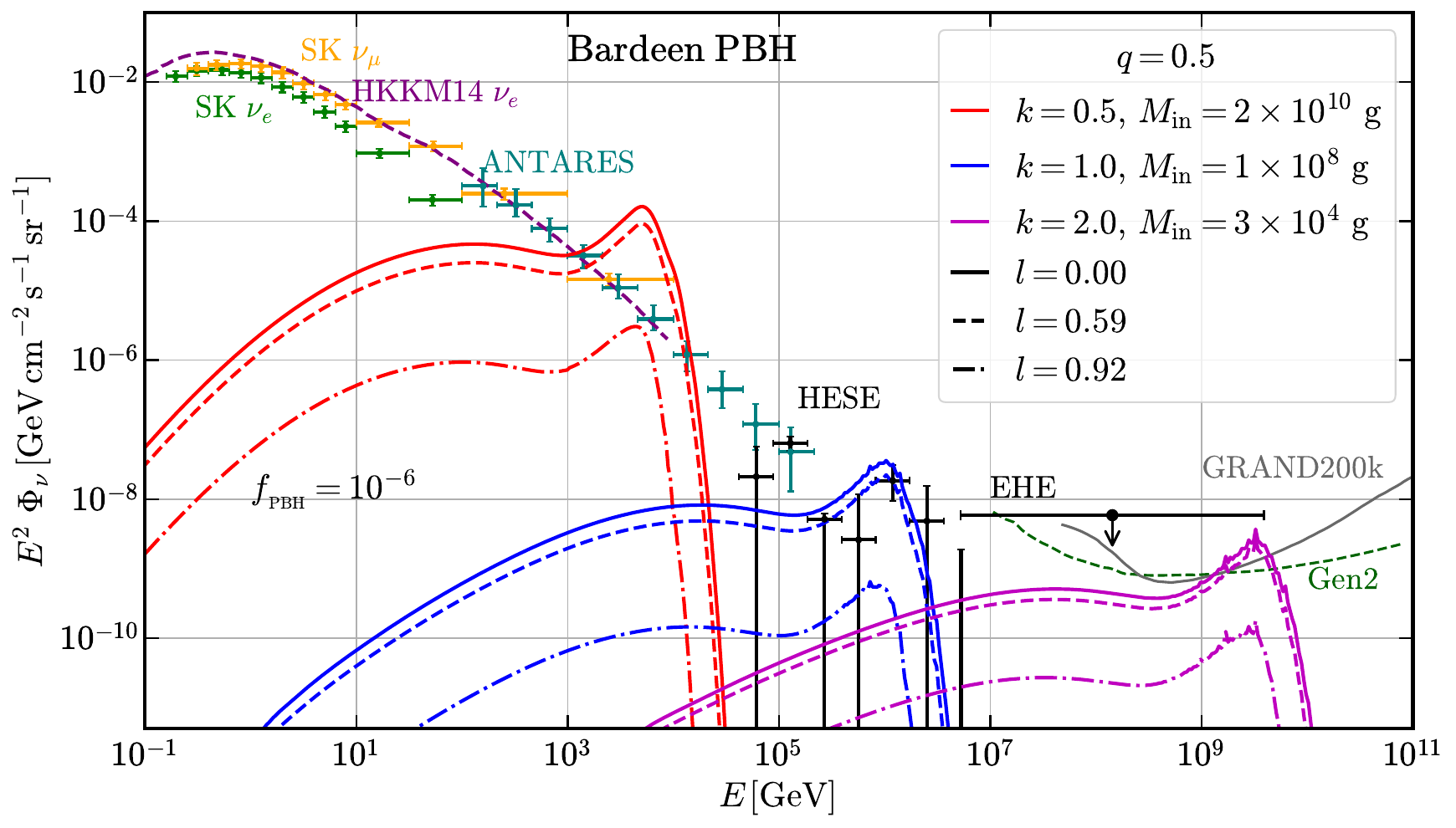}
    \includegraphics[width=.9\linewidth]{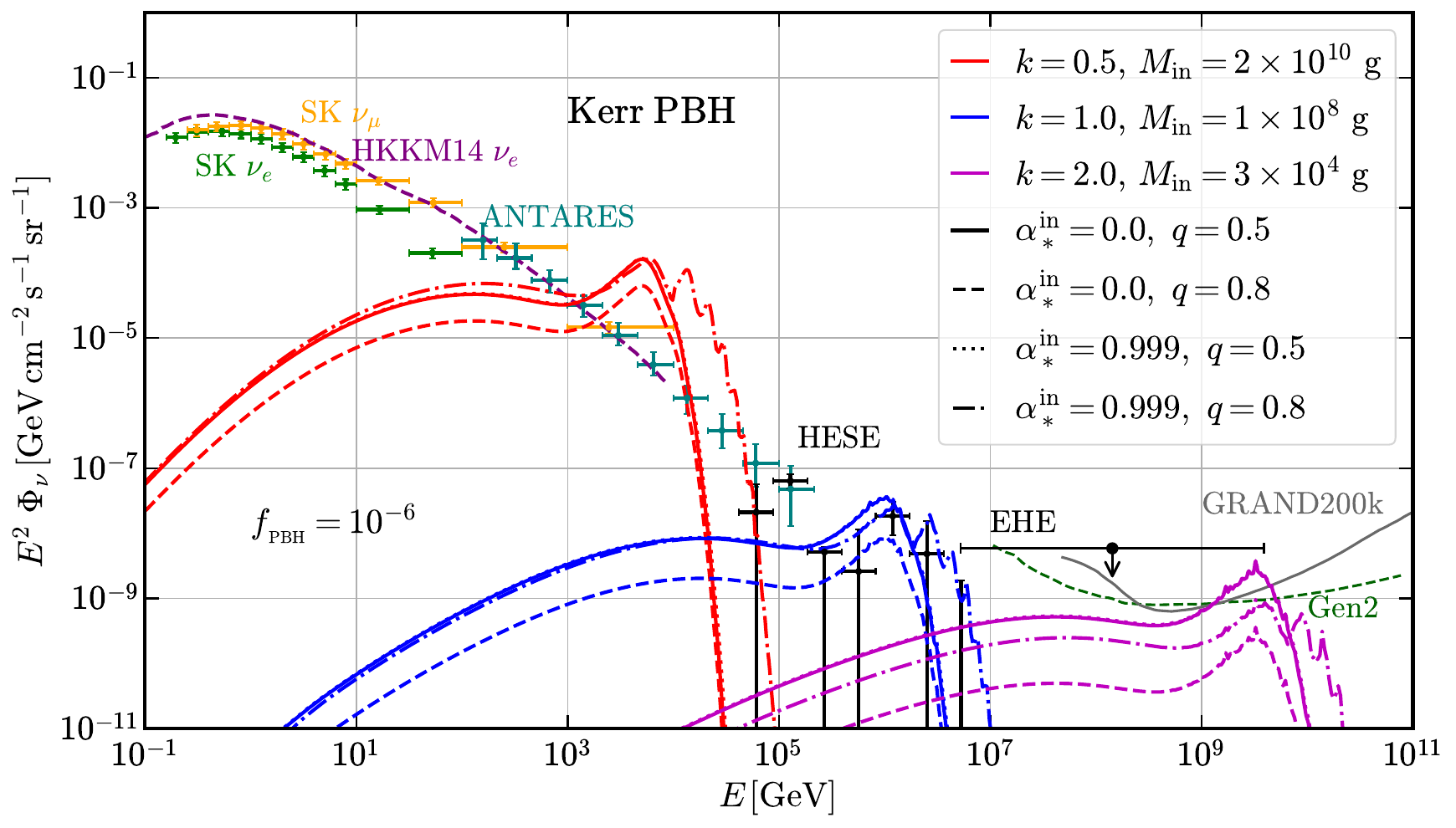}
    \caption{The total neutrino fluxes (summed over all neutrino flavours) from the evaporation of memory burdened Bardeen PBHs (top panel) and Kerr PBHs (bottom panel) are shown as a function of neutrino energy. The different colours and line styles correspond to various combinations of PBH mass, Bardeen parameter, spin, and memory burden parameter, as indicated in the figure.
    The quantity $\fpbh$ is fixed at $10^{-6}$.
    The observations are listed as $7.5$-year HESE data~\cite{IceCube:2020wum},
 $7$-years EHE neutrinos upper bound~\cite{IceCube:2016uab},
 Grand200k~\cite{GRAND:2018iaj},
 IceCube-Gen2~\cite{IceCube:2019pna,IceCube-Gen2:2020qha}, SK~\cite{Super-Kamiokande:2015qek}, HKKM14~\cite{Honda:2015fha,Hyper-Kamiokande:2018ofw}, and
ANTARES~\cite{ANTARES:2013iuz}.}
    \label{fig:flux-demo}
\end{figure*}

\section{High-energy neutrino flux from PBHs}\label{sec:nutrino-flux}

In this section, we discuss the high-energy neutrino flux produced by PBH evaporation. Throughout this work, we assume a monochromatic PBH mass function. Let $N_{\nu_i}$ denote the number of neutrinos of flavor $i$ emitted through the Hawking evaporation of a single PBH, where $\nu_i$ represents the sum of neutrinos and antineutrinos of that flavor. The differential primary neutrino emission rate per unit time and unit energy is given by
\begin{align}
    \f{\d^2N_{\nu_i}^{\rm pri}}{\d E\,\d t}
    &=
    \f{g_{\nu_i}}{2\pi}
    \sum_{\tilde{\ell},m}
    \f{\Gamma_{\tilde{\ell}m}^{(1/2)}(E)}
    {\exp\!\left[(E-m\Omega_{\rm H})/\Tbh\right]+1},
    \label{eq:dndtdw1}
\end{align}
where $g_{\nu_i}=2$ denotes the degrees of freedom of one neutrino flavor, including the corresponding antineutrino. Therefore, summing over the three flavors gives a total of six neutrino degrees of freedom. The quantities $\tilde{\ell}$ and $m$ are the angular-mode quantum numbers, while $\Gamma_{\tilde{\ell}m}^{(1/2)}(E)$ is the corresponding spin-$1/2$ greybody factor. We use $\tilde{\ell}$ to distinguish the angular quantum number from the regularization scale $\ell$ and the dimensionless Bardeen parameter $l$.

For a Kerr PBH, the angular velocity of the event horizon is
\begin{align}
    \Omega_{\rm H}
    &=
    \f{a}{r_+^2+a^2}
    =
    \f{4\pi\Mpl^2}{M}
    \f{\alpha_{\ast}}
    {1+\sqrt{1-\alpha_{\ast}^2}}.
\end{align}
Thus, the combination $E-m\Omega_{\rm H}$ in Eq.~\eqref{eq:dndtdw1} accounts for the rotational chemical potential in the Kerr Hawking spectrum. Retaining this term is important when calculating the emission from spinning PBHs. For the nonrotating Bardeen and Schwarzschild geometries, $\Omega_{\rm H}=0$, and Eq.~\eqref{eq:dndtdw1} reduces to the usual Fermi--Dirac spectrum with the denominator $\exp(E/\Tbh)+1$.

We calculate the neutrino emission spectra numerically using the code \texttt{BlackHawk}~\cite{Arbey:2019mbc,Arbey:2021mbl,Arbey:2026koc}. We also include hadronization and particle-decay processes through \texttt{HDMSpectra} to calculate the secondary neutrino contribution~\cite{Bauer:2020jay}. The total neutrino spectrum emitted by a PBH is therefore
\begin{align}
    \f{\d^2N_{\nu_i}}{\d E\,\d t}=
    \f{\d^2N_{\nu_i}^{\rm pri}}{\d E\,\d t}+
    \f{\d^2N_{\nu_i}^{\rm sec}}{\d E\,\d t}.
\end{align}

The memory burden suppresses PBH evaporation and consequently reduces the neutrino emission rate. To describe both the semiclassical and memory burdened phases within a single expression, we write
\begin{align}
    \f{\d^2N_{\nu_i}^{\rm mb}}{\d E\,\d t}
    =
    S^{-k\Theta(q\Min-M)}
    \f{\d^2N_{\nu_i}}{\d E\,\d t}.
\end{align}
Before the PBH mass reaches $q\Min$, the Heaviside function vanishes and the standard Hawking emission spectrum is recovered. Once $M\leq q\Min$, the emission rate is suppressed by the factor $S(M)^{-k}$.

The total high-energy neutrino flux receives contributions from both the Galactic DM halo and the approximately isotropic extragalactic PBH distribution. Summing over the three neutrino flavors, the total flux is given by
\begin{align}
    \Phi_\nu(E)
    =
    \sum_{i=1}^{3}
    \left[
        \f{\d^2\phi_{\nu_i}^{\rm gal}}
        {\d E\,\d\Omega}
        +
        \f{\d^2\phi_{\nu_i}^{\rm egal}}
        {\d E\,\d\Omega}
    \right].
\end{align}

We define the present PBH mass as $M_0\equiv M(t_0)$. For long-lived PBHs whose mass changes negligibly during the memory burdened phase, one may use the approximation $M_0\simeq q\Min$. The Galactic contribution, averaged over the relevant solid angle, is then given by
\begin{align}
    \f{\d^2\phi_{\nu_i}^{\rm gal}}
    {\d E\,\d\Omega}
    =
    \f{\fpbh\,\cJ}{4\pi M_0}
    \f{\d^2N_{\nu_i}^{\rm mb}}
    {\d E\,\d t}
    \!\left(E;M_0,\alpha_{\ast}(t_0),l\right).
\end{align}
Here, $\fpbh$ denotes the present fraction of DM in PBHs. The sky-averaged line-of-sight factor is defined as
\begin{align}
    \cJ \equiv
    \f{1}{4\pi}
    \int\d\Omega
    \int_{\rm l.o.s.}\d s\,
    \rho_{\rm NFW}\!\left[r(s,\Omega)\right].
\end{align}
We use $\cJ=2.22\times10^{22}~{\rm GeV\,cm^{-2}\,sr^{-1}}$, obtained from an NFW DM density profile with a scale radius of $25~{\rm kpc}$ and a local DM density of $0.4~{\rm GeV\,cm^{-3}}$~\cite{Iocco:2015xga,Benito:2019ngh,Benito:2020lgu}. The ${\rm sr}^{-1}$ in the quoted units emphasizes the convention that $\cJ$ is averaged over the solid angle; steradians are dimensionless in the corresponding dimensional analysis.
The extragalactic contribution is given by
\begin{align}
    \f{\d^2\phi_{\nu_i}^{\rm egal}}
    {\d E\,\d\Omega}
    &=
    \f{\fpbh\,\rho_{\rm DM}}{4\pi M_0}
    \int_{t_{\rm eq}}^{t_0}
    \d t\,[1+z(t)]
    \nonumber\\
    &\quad\times
    \f{\d^2N_{\nu_i}^{\rm mb}}
    {\d E\,\d t}
    \!\left(E_z;M(t),\alpha_{\ast}(t),l\right).
\end{align}
where $\rho_{\rm DM}=1.26\times10^{-6}~{\rm GeV\,cm^{-3}}$ is the present cosmological DM energy density~\cite{Planck:2018vyg}. The emission spectrum is evaluated at the redshifted energy
\begin{align}
    E_z=E\,[1+z(t)],
\end{align}
where $E$ is the neutrino energy measured at Earth. The factor $1+z(t)$ in the time integral accounts for the relation between the emitted and observed energy intervals. The time dependence of the PBH mass and, for Kerr PBHs, its spin is included in the source spectrum.

Note that the integration is performed from matter--radiation equality, $t_{\rm eq}$, to the present age of the Universe, $t_0$. Emission before matter--radiation equality is neglected because it is more strongly redshifted and is assumed to provide a subdominant contribution in the high-energy range considered here.

In Fig.~\ref{fig:flux-demo}, we plot the total neutrino flux, including both Galactic and extragalactic contributions and summed over all neutrino flavors, as a function of the neutrino energy. The results for memory burdened Bardeen and Kerr PBHs are shown in the top and bottom panels, respectively. In both panels, the different colors correspond to different combinations of the memory burden parameter $k$ and the initial PBH mass $\Min$. The line styles represent different values of the Bardeen parameter $l$ in the top panel and different combinations of the initial spin $\alpha_{\ast}^{\rm in}$ and $q$ in the bottom panel, as indicated in the legends. The PBH abundance is fixed at $\fpbh=10^{-6}$.

The neutrino spectrum contains primary and secondary components. Secondary neutrinos produced through hadronization and particle decays are particularly important at energies below the characteristic Hawking scale, whereas the primary component becomes increasingly important near the spectral peak and toward the high-energy endpoint. For Bardeen PBHs, the predicted neutrino flux varies significantly with $l$: increasing the Bardeen parameter suppresses the flux owing to the corresponding reduction in the Hawking temperature and evaporation rate. For Kerr PBHs, the effect of spin is particularly noticeable for the larger value $q=0.8$. In this case, the memory burdened phase begins before substantial spin-down occurs, so that the residual spin $\alpha_{\ast}^{\rm q}$ remains significant and produces a clear difference between the fluxes from spinning and nonspinning PBHs.

For comparison, we also show the $7.5$-year IceCube high-energy starting event (HESE) data~\cite{IceCube:2020wum}, the upper bound on extremely high-energy (EHE) neutrinos obtained from seven years of IceCube data~\cite{IceCube:2016uab}, the projected sensitivities of GRAND200k~\cite{GRAND:2018iaj} and IceCube-Gen2~\cite{IceCube:2019pna,IceCube-Gen2:2020qha}, the Super-Kamiokande data~\cite{Super-Kamiokande:2015qek}, the Hyper-Kamiokande sensitivity based on the HKKM14 atmospheric-neutrino flux~\cite{Honda:2015fha,Hyper-Kamiokande:2018ofw}, and the ANTARES data~\cite{ANTARES:2013iuz}. 

\section{Details of statistical analysis}\label{sec:stat-analysis}

\begin{figure}
    \centering
    \includegraphics[width=\linewidth]{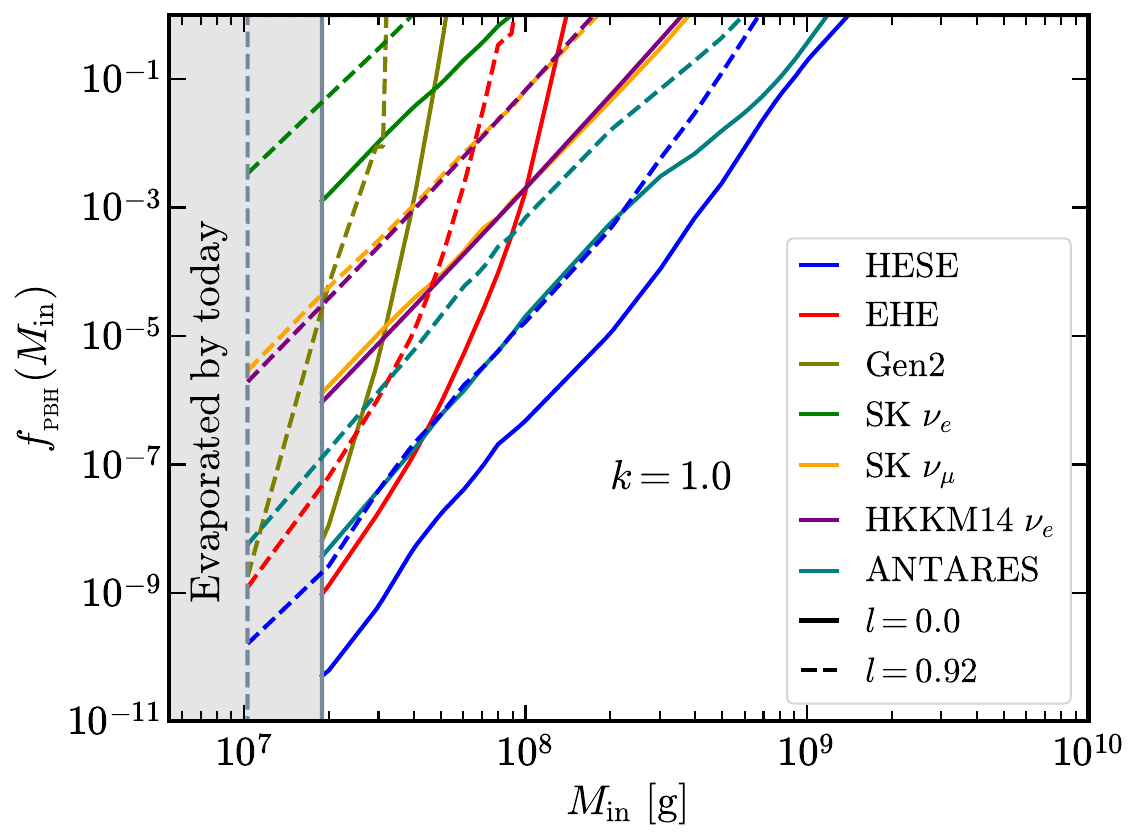}
    \includegraphics[width=\linewidth]{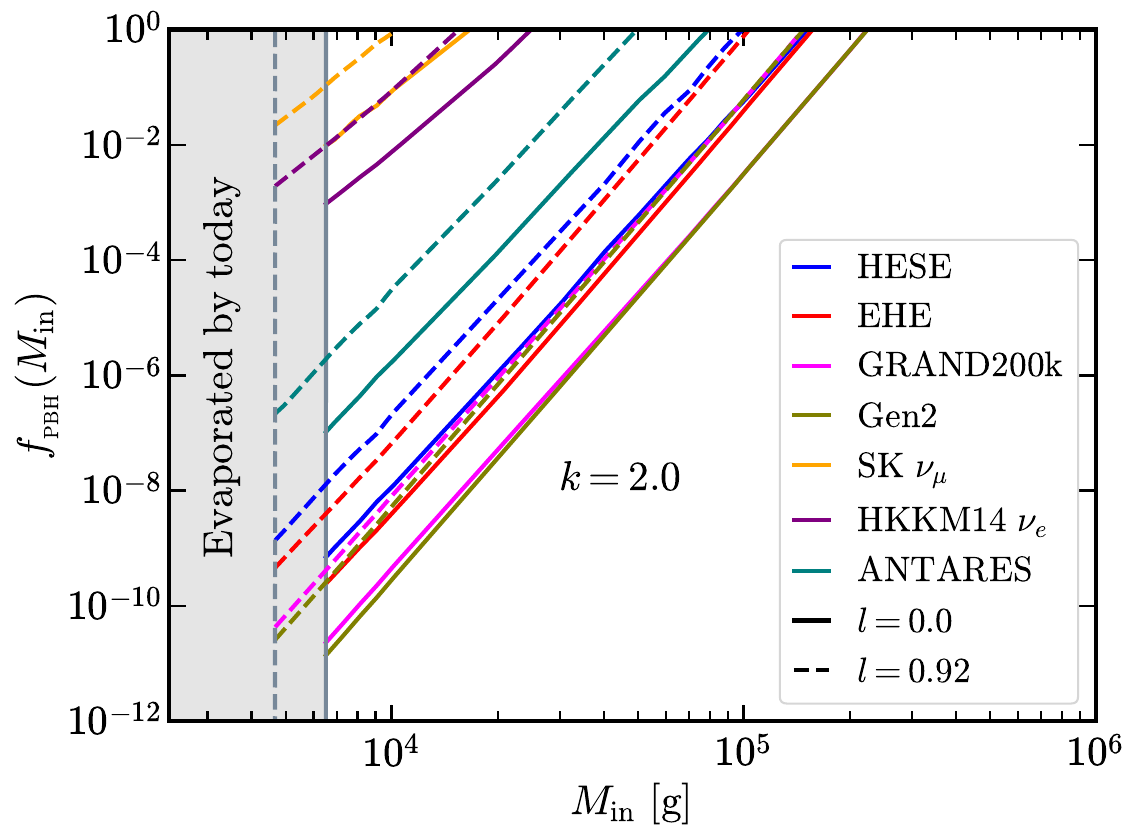}
    \includegraphics[width=\linewidth]{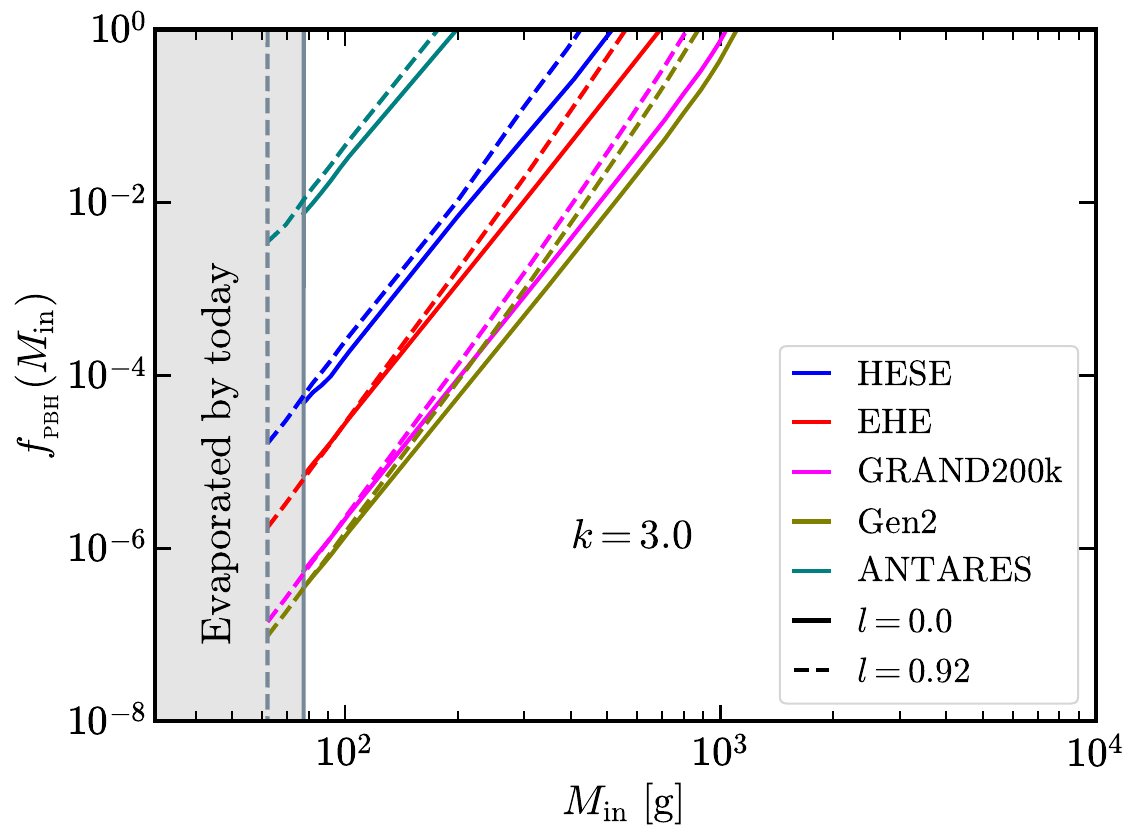}

    \caption{Constraints on $\fpbh$ derived from various neutrino observations are shown as a function of the PBH mass for Bardeen PBHs. The case of $k=1$, $2$, and $3$ are shown in top, middle and bottom panel, respectively and we have taken $q=1/2$.
    The solid and dashed curves represent PBHs with Bardeen parameters $l=0.0$ and $l=0.92$, respectively. The grey lines indicate the maximum value of $\Min$ below which the PBHs are evaporated by the present time.}
    \label{fig:constraints-bardeen}
\end{figure}

\begin{figure*}
    \centering
    \includegraphics[width=0.49\linewidth]{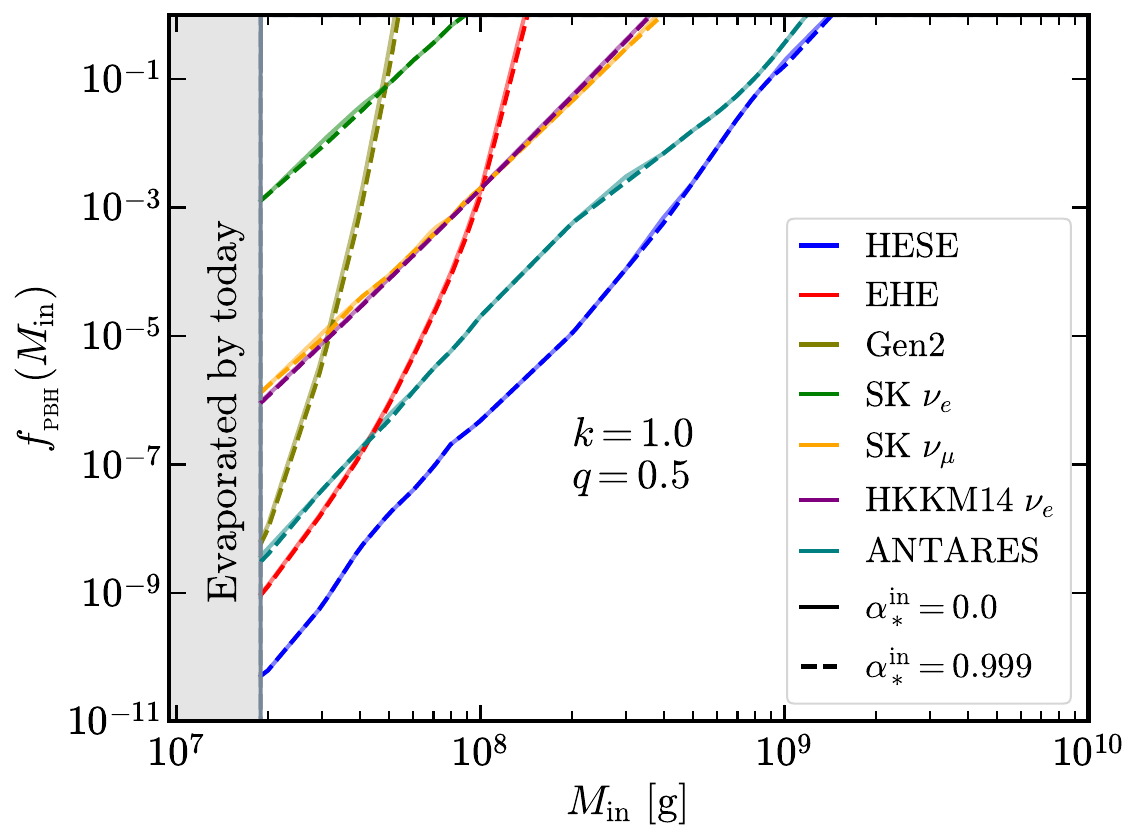}
    \includegraphics[width=0.49\linewidth]{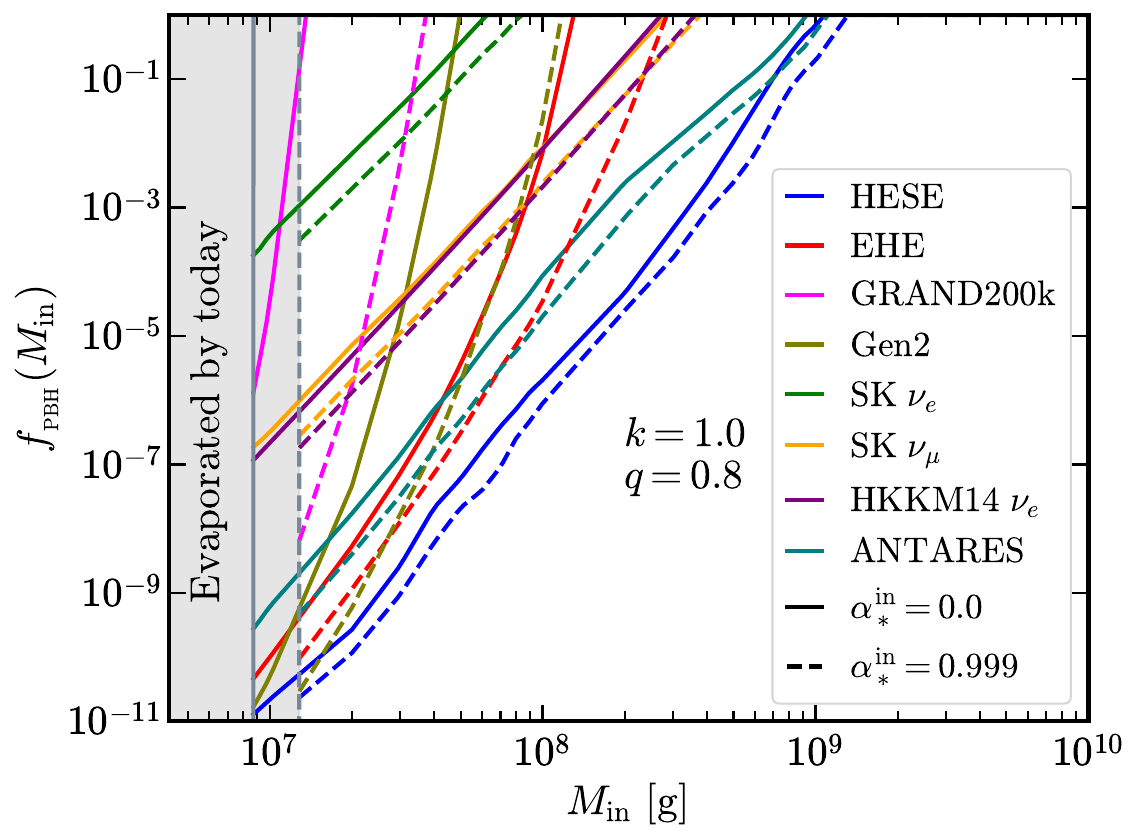}
    \includegraphics[width=0.49\linewidth]{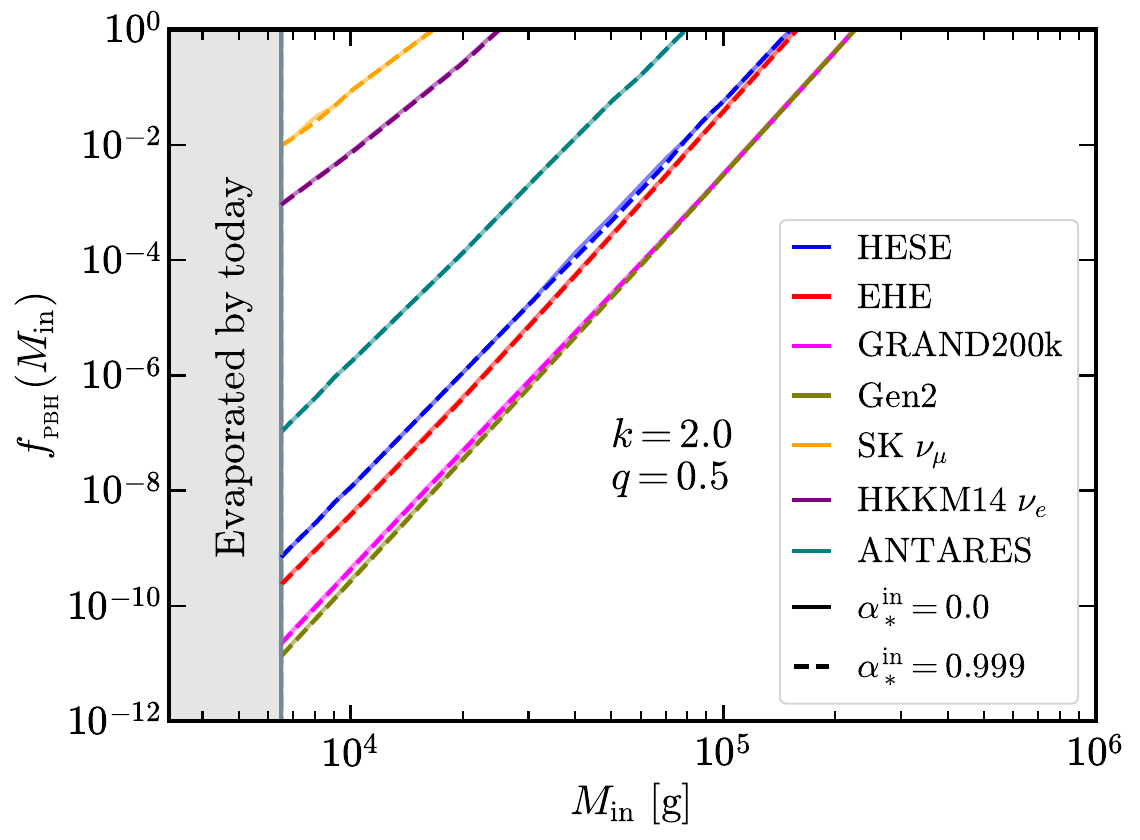}
    \includegraphics[width=0.49\linewidth]{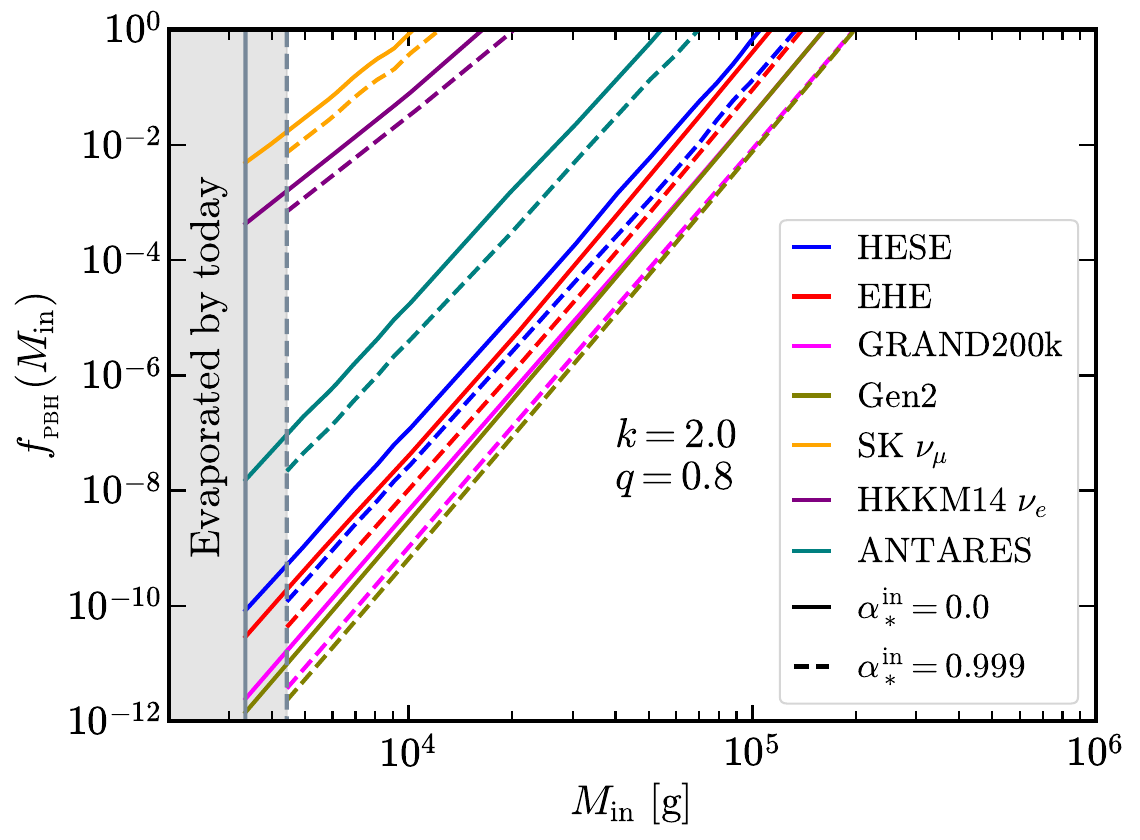}
    \includegraphics[width=0.49\linewidth]{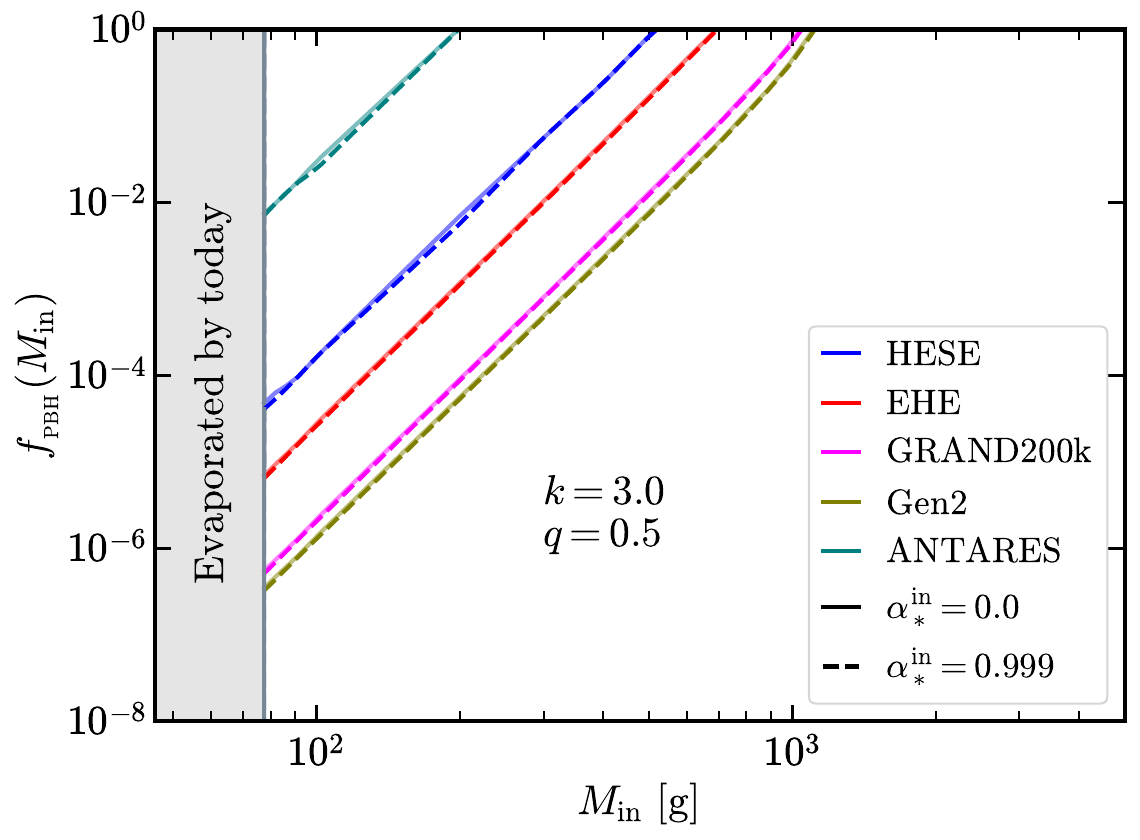}
    \includegraphics[width=0.49\linewidth]{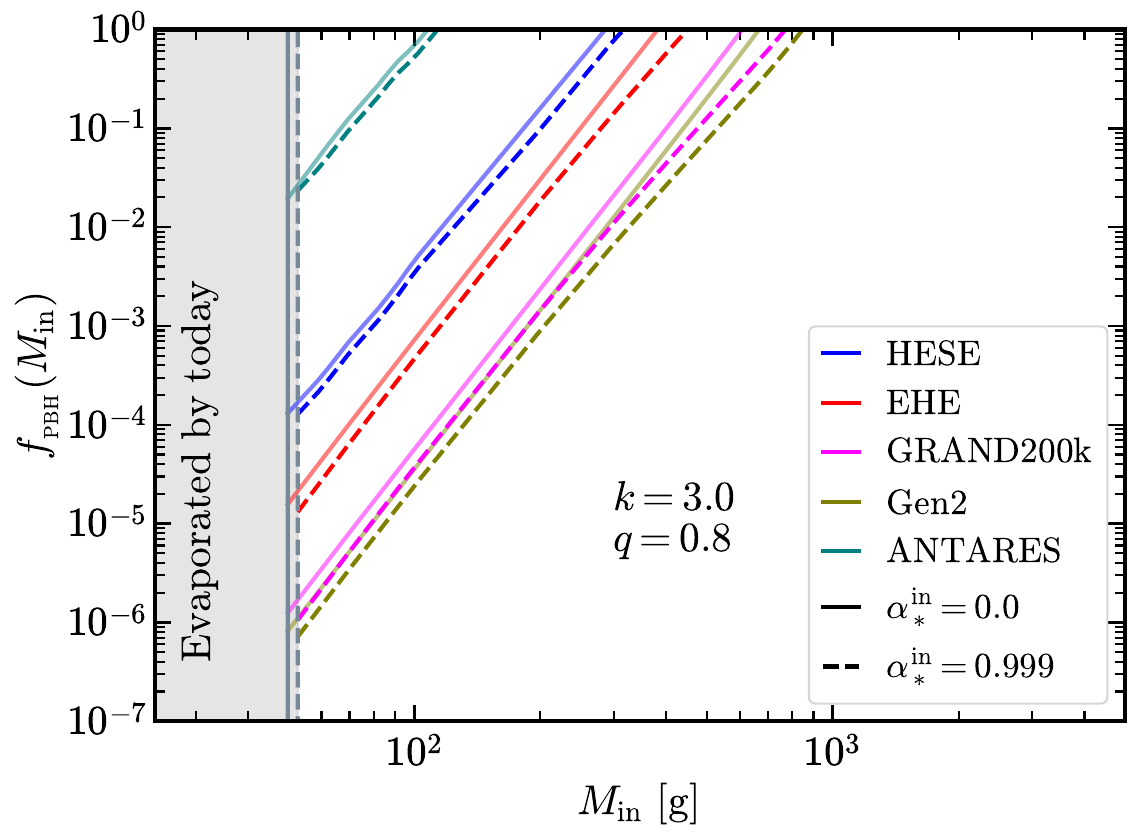}
    \caption{Constraints on $\fpbh$ derived from various neutrino observations are shown as a function of the PBH mass. The top, middle, and bottom panels correspond to the memory burden parameter values $k=1$, $2$, and $3$, respectively. The left and right panels show the cases $q=0.5$ and $q=0.8$, respectively. Solid and dashed curves represent PBHs with initial spin parameters $\alpha_\ast^{\rm in}=0.0$ and $\alpha_\ast^{\rm in}=0.999$, respectively. The grey lines indicate the maximum value of $\Min$ below which the PBHs are evaporated by the present time.}
    \label{fig:constraints-kerr}
\end{figure*}

To derive constraints on $\fpbh$, we adopt two complementary statistical methods. When observational data are available, we employ the following one-sided, background-agnostic likelihood function:
\begin{align}
    \cL(\fpbh,\Min,k)
    =
    \prod_{i=1}^{n_{\rm data}}
    \begin{cases}
        \widetilde{\cP}(d_i\vert\mu_i),
        & \mu_i>d_i,\\
        1,
        & \mu_i\leq d_i,
    \end{cases}
\end{align}
where $d_i$ is the observed value at the $i$th data point and $\mu_i(\Min,\fpbh,k)$ is the corresponding prediction, including the PBH contribution. Here, $\widetilde{\cP}$ denotes the probability distribution normalized to its maximum value, such that $\widetilde{\cP}(d_i\vert d_i)=1$. This prescription ensures that predictions below the observed flux are not penalized. The index $i$ runs over all data points included in the analysis. The dependence on additional parameters, such as $l$, $q$, and $\alpha_{\ast}^{\rm in}$, is left implicit because these parameters are fixed separately for each analysis. We define
\begin{align}
    \Delta\chi^2=-2\ln\cL
\end{align}
and derive the upper limit on $\fpbh$ using the appropriate critical value for one degree of freedom under the Wilks approximation.

\begin{itemize}

\item
For IceCube-HESE, SK, and ANTARES, we take $\widetilde{\cP}$ to be a normalized Gaussian likelihood. For the HESE analysis, we use the frequentist flux measurements as the observational data.

\item
For the seven-year IceCube EHE data, we adopt a Poisson likelihood for the number of events observed within the energy interval $[E_{\rm min},E_{\rm max}]$. In the energy interval considered in our analysis, we take the observed event count to be one. The expected number of events generated by PBH evaporation is
\begin{align}
    n_{\rm events}
    =
    4\pi T_{\rm obs}
    \int_{E_{\rm min}}^{E_{\rm max}}
    \d E\,
    \Phi_\nu(E)A_{\rm eff}(E),
    \label{eq:nevents}
\end{align}
where $\Phi_\nu(E)$ is the all-flavor neutrino flux generated by PBH evaporation, $A_{\rm eff}(E)$ is the corresponding all-flavor effective area of the detector, and $T_{\rm obs}$ is the observation time. For the seven-year IceCube EHE analysis, we take $T_{\rm obs}=7~{\rm yr}$. The factor $4\pi$ accounts for the full-sky solid angle and assumes an isotropic neutrino flux.

\item
For future neutrino observatories, including IceCube-Gen2, GRAND200k, and Hyper-Kamiokande, we assume that no events are observed within the relevant energy range over an exposure time of three years. Assuming a negligible background, we derive the 95\% confidence-level upper limit on $\fpbh$ by requiring
    $n_{\rm events}<3.09,$
following the Feldman-Cousins prescription for zero observed events and zero expected background~\cite{Feldman:1997qc}.
We infer the detector effective area from the published sensitivity curves following the procedure described in Refs.~\cite{ARA:2015wxq,Chianese:2021htv}. When $S(E)$ denotes the differential sensitivity per unit energy, area, time, and solid angle, the effective area is given by
\begin{align}
    A_{\rm eff}(E)
    =
    \f{2.44}
    {4\pi\ln(10)\,E\,S(E)\,T_{\rm obs}}.
\end{align}
Here, $2.44$ is the Feldman-Cousins upper limit at 90\% C.L. for zero observed events and zero expected background, while the factor $\ln(10)$ arises from integration over a logarithmic energy bin of one decade. This conversion therefore applies when the published curve represents a 90\% C.L. differential sensitivity. After reconstructing $A_{\rm eff}(E)$, we apply the 95\% C.L. condition $n_{\rm events}<3.09$ to derive the projected constraint on $\fpbh$.
If the published sensitivity is instead presented in the commonly used form $\mathcal{S}(E)=E^2S(E)$, the equivalent expression is
\begin{align}
    A_{\rm eff}(E)
    =
    \f{2.44\,E}
    {4\pi\ln(10)\,\mathcal{S}(E)\,T_{\rm obs}}.
\end{align}
Note that in the context of neutrino constraints on nonspinning Schwarzschild PBHs, the assumption of zero observed events represents a plausible scenario for future neutrino observations. We adopt the same assumption when deriving the projected constraints on Bardeen and Kerr PBHs.

\end{itemize}

\section{Results and discussions}\label{sec: result}

The main goal of our analysis is to constrain the abundance of memory burdened Bardeen and Kerr PBHs using current neutrino observations and the projected sensitivities of future experiments. The Bardeen regularization parameter and Kerr rotation modify the evaporation history and neutrino spectrum in different ways, leading to distinct constraints relative to the Schwarzschild case.

We first consider the case of Bardeen PBHs. In Fig.~\ref{fig:constraints-bardeen}, we present the resulting upper limits on $\fpbh$ as a function of the initial PBH mass $\Min$. For Bardeen PBHs, increasing $l$ at fixed $k$ and $\Min$ suppresses the Hawking temperature and the total evaporation rate. This reduces the neutrino-flux amplitude and suppresses the high-energy tail, narrowing the energy range over which the flux is appreciable. Consequently, the limits on $\fpbh$ are weaker for $l=0.92$ than for the Schwarzschild limit $l=0$.

We next consider spinning PBHs described by the Kerr geometry. In Fig.~\ref{fig:constraints-kerr}, we present the corresponding upper limits on $\fpbh$ as a function of $\Min$. For Kerr PBHs, the effect of the initial spin depends strongly on the value of $q$. For $q=0.8$, rapidly spinning PBHs yield substantially stronger constraints than nonspinning PBHs. In this case, the memory burdened phase begins before substantial spin-down occurs, leaving a significant residual spin $\alpha_{\ast}^{\rm q}$. The rotational contribution $m\Omega_{\rm H}$ and the spin-dependent greybody factors consequently enhance and broaden the neutrino spectrum. For $q=0.5$, more angular momentum is radiated before the onset of the memory burdened phase, and the difference between the spinning and nonspinning constraints is correspondingly smaller.

All upper limits shown in Figs.~\ref{fig:constraints-bardeen} and~\ref{fig:constraints-kerr} are presented at the $95\%$ confidence level. For each observational data set, the parameter space above the corresponding curve is excluded. As a consistency check, for $k=2$ and $q=0.5$, the Bardeen result with $l=0$ and the Kerr result with $\alpha_{\ast}^{\rm in}=0$ both reduce to the Schwarzschild case. The corresponding constraints obtained from HESE, EHE, IceCube-Gen2, and GRAND200k agree with those reported in Ref.~\cite{Chianese:2024rsn}.

The experiment providing the strongest constraint depends on both $\Min$ and $k$. For $k=1$, the dominant limits arise from HESE, whereas IceCube-Gen2 and GRAND200k provide the strongest sensitivities for $k=2$ and $3$. At fixed geometry and PBH mass, increasing $k$ suppresses the flux normalization through the energy-independent factor $S^{-k}$ without shifting the spectral peak. A larger $k$, however, allows lighter PBHs to survive until the present epoch. Since the characteristic Hawking energy scales approximately as $T_{\rm BH}\propto M^{-1}$, the relevant spectra move toward higher energies. Thus, the reduction in amplitude follows directly from the stronger memory suppression, while the shift toward higher energies reflects the smaller masses of the surviving PBHs.

Overall, the memory burden and geometric parameters play complementary roles. The parameters $k$ and $q$ primarily determine the PBH lifetime and the surviving mass range, whereas $l$ and $\alpha_{\ast}^{\rm q}$ control the temperature, greybody factors, and spectral properties of the present-day emission. Bardeen regularization weakens the neutrino constraints by suppressing the emission, while a sufficiently large residual Kerr spin can strengthen them. Current and future neutrino experiments therefore provide complementary probes of memory burdened PBHs and their possible contribution to the DM abundance.
\begin{acknowledgments}
The authors thank Fabio Iocco and S.~K.~Jeesun for useful discussions. M.R.H. acknowledges the Tsung-Dao Lee Institute at Shanghai Jiao Tong University for financial support through the Siyuan Postdoctoral Fellowship.
\end{acknowledgments}

\bibliographystyle{apsrev4-1}
\bibliography{references}

\end{document}